\documentclass[onecolumn,aps,prl,preprint,superscriptaddress,footinbib,amsmath,amssymb,aps,floatfix]{revtex4-1}

\usepackage{bm}
\usepackage{graphicx}
\usepackage{setspace}

\newcommand{\ket}[1]{\left|#1\right\rangle}

\newcommand{\kB}{\ensuremath{k_\mathrm{B}}}
\renewcommand{\vec}[1]{\mathbf{#1}}
 
\newcommand{\us}{\ensuremath{\mathrm{\mu s}}}
\newcommand{\pdagger}{{\phantom{\dagger}}}
\newcommand{\dd}{\mathrm{d}}
\newcommand{\ff}{\mathrm{f}}

\begin{document}

\title{Evidence for universal relations describing a gas with $p$-wave interactions}
\author{Christopher Luciuk}
\author{Stefan Trotzky}
\author{Scott Smale}
\affiliation{Department of Physics, University of Toronto, M5S 1A7 Canada}
\author{Zhenhua Yu}
\affiliation{Institute for Advanced Study, Tsinghua University, Beijing 100084, China}
\author{Shizhong Zhang}
\email{shizhong@hku.hk}
\affiliation{Department of Physics, Centre of Theoretical and Computational Physics, University of Hong Kong, China}
\author{Joseph H.\ Thywissen}
\email{joseph.thywissen@utoronto.ca}
\affiliation{Department of Physics, University of Toronto, M5S 1A7 Canada}
\affiliation{Canadian Institute for Advanced Research, Toronto, M5G 1Z8 Canada}

\begin{abstract}
{\bf
Thermodynamics provides powerful constraints on physical and chemical systems in equilibrium. However, non-equilibrium dynamics depends explicitly on microscopic properties, requiring an understanding beyond thermodynamics. Remarkably, in dilute gases, a set of universal relations is known to connect thermodynamics directly with microscopic properties. So far, these ``contact'' relations have been established only for interactions with $\bm{s}$-wave symmetry, i.e., without relative angular momentum. We report measurements of two new physical quantities, the ``$\bm{p}$-wave contacts'', and present evidence that they encode the universal aspects of $\bm{p}$-wave interactions through recently proposed relations. Our experiments use an ultracold Fermi gas of $\bm{^{40}}$K, in which $\bm{s}$-wave interactions are suppressed by polarising the sample, while $\bm{p}$-wave interactions are enhanced by working near a scattering resonance. Using time-resolved spectroscopy, we study how correlations in the system develop after ``quenching'' the atoms into an interacting state. Combining quasi-steady-state measurements with new contact relations, we infer an attractive $\bm{p}$-wave interaction energy as large as half the Fermi energy. Our results reveal new ways to understand and characterise the properties of a resonant $\bm{p}$-wave quantum gas.}
\end{abstract}

\maketitle

A fundamental question provoked by observation of natural systems is how macroscopic and collective properties depend on microscopic few-body interactions. Ultracold neutral atoms provide a model system in which to explore this question, since in certain conditions, few-body interactions can be tuned and characterised precisely. Over the last decade, a direct link has been made in these systems between thermodynamic properties and the underlying isotropic ($s$-wave) interactions. At the centre stage is a quantity called the ``contact''~\cite{Tan2008:01,Tan2008:02,Tan2008:03,Werner2009,Zhang2009,Braaten:2012gh}, which describes how the energy of a system changes when the interaction strength is changed. Surprisingly, the contact is also the pivot of a set of universal relations, that constrain numerous {\em microscopic} properties, including the two-particle correlation function at short range. These relations apply regardless of temperature, density, or interaction strength~\cite{Tan2008:01,Tan2008:02,Tan2008:03,Zhang2009}, to fermions and bosons~\cite{Werner2012,Werner:2012hy,Wild:2012fi}, and in one-, two-, and three-dimensional systems~\cite{Olshanii:2003kn,Combescot:2009gw,Frohlich:2012ic,Werner2012}. Contact relations have also been extended to Coulomb gases~\cite{Barth:2011ky} and neutron-proton interactions~\cite{Weiss:2015gf}. Despite the breadth of this discussion, measurements of the contact have so far been restricted to systems with $s$-wave interactions.

\begin{figure}[p] 
\begin{center} \includegraphics[width=0.75\textwidth]{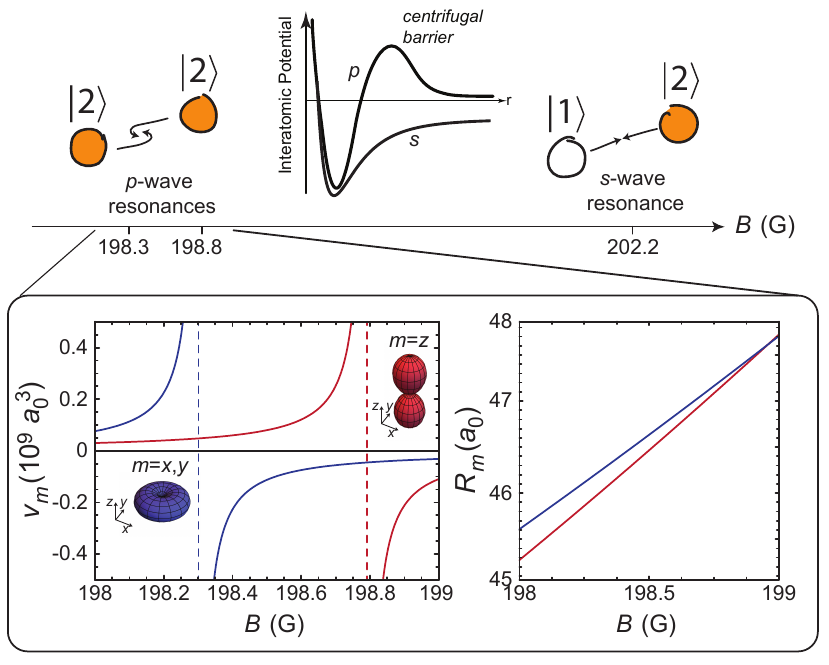} \end{center}
\caption{ {\bf Tuning $\bm{p}$-wave interactions in $^{\bm{40}}$K.} In a trapped gas of fermionic $^{40}$K, the nature of interactions can be controlled by selecting internal states and tuning the magnetic field. Near the $p$-wave resonance studied here, {\em spin-triplet} collisions are resonant for atoms in the $\ket{2}$ state. This can be contrasted with the $s$-wave interactions near a Feshbach resonance at 202.2 G, where {\em spin-singlet} collisions in a $\ket{1}+\ket{2}$ mixture are enhanced. Unlike $s$-waves, $p$-waves are normally suppressed at low energy due to a centrifugal energy barrier. The scattering volume $v_m$ diverges at distinct $m=z$ and $m={x,y}$ Feshbach resonances, while $R_m$ is only weakly dependent on $B$ and nearly isotropic. A parameterization of $v_m$ and $R_m$ is given in Methods.}
\label{fig:cartoon}
\end{figure}

In general, the relative wave function of any pair of particles can be decomposed into components with angular momentum equal to an integer multiple $\ell$ of $\hbar$ quanta. In a spin-polarised Fermi gas, quantum statistics forbids short-range interactions with even values of $\ell$. Therefore, the first allowed scattering channel has $\ell=1$ ($p$-wave), which is typically weak due to the centrifugal barrier (see Fig.~\ref{fig:cartoon}): the scattering cross section decreases with the square of the collision energy~\cite{DeMarco1999}. However, resonant enhancement of $p$-wave collisions has been observed in $^{40}$K and $^6$Li~\cite{Regal2003,Zhang2004,Schunck:2005cf}, raising the possibility of studying a gas with strong, tuneable $p$-wave interactions.

Interest in $p$-wave systems originated with liquid $^3$He, which at low temperature is a superfluid of pairs with $\ell=1$ in a spin triplet state. It is thought that Sr$_2$RuO$_4$ realises a chiral $p_x + i p_y$ superconductor, although definitive evidence is still elusive~\cite{Kallin:2012kx}. An ultracold gas with tuneable $p$-wave interactions could be employed to explore the evolution from weak- to strong-coupling superfluids, including the topological quantum phase transition predicted in two-dimensional samples~\cite{Read:2000iq,Levinsen:2007gy}. In certain conditions, a $p$-wave superfluid could host Majorana modes that exhibit non-Abelian statistics, which is important for topological quantum computing~\cite{Kallin:2012kx,MajoranaReview}.

Here we report on time-resolved spectroscopic characterisation of an ultracold spin-polarised Fermi gas with near-resonant $p$-wave interactions. We find that both the momentum distribution and the radio-frequency (rf) response follow asymptotic scaling consistent with $p$-wave contact relations \cite{Inotani2012,Yoshida2015,Yu2015,Zhou:2016}, Furthermore, contact values measured with these two independent methods show good agreement. Then, relying upon the validity of the complete contact theory, we interpret the dynamical evolution of the contact as the population dynamics in the closed channel.

\bigskip
{\bf Near-resonant $\bm{p}$-wave interactions}

For $p$-wave scattering, the van der Waals interaction between atoms can be regarded as short-ranged~\cite{Zhang2010}: i.e., acting only when $r < r_0$, where $r$ is the inter-nuclear separation and $r_0$ is comparable to the van der Waals length, about $3$\,nm for potassium. In ultracold Fermi gases, the inter-particle separation is much larger: $1/k_\mathrm{F}$ is typically $10^2$\,nm, where $\hbar k_\mathrm{F}$ is the Fermi momentum. As a result, the low-energy interaction can be described completely by the $p$-wave scattering phase-shift $\delta_{1m}$, $k^3\cot\delta_{1m}(k)=-1/v_m-k^2/R_m$, where $k$ is the relative wave vector, $v_m$ is the scattering volume, $R_m$ is the effective range, and $m=\{x,y,z\}$ labels the projection of angular momentum onto the magnetic-field axis~\cite{Ticknor2004,Zhang2010}.

Figure~\ref{fig:cartoon} depicts the control introduced by a Feshbach resonance \cite{Chin2010}, where a closed-channel molecular bound state $\psi_c$ is tuned near energetic resonance with an open-channel scattering state $\psi_o$ of a pair of free atoms. Tuning is accomplished with a magnetic field due to a differential magnetic moment $\delta \mu$ between the closed and open channels. Far from resonance, $v_m$ and $R_m$ take on background values $v^{\rm bg}_m$ and $R^{\rm bg}_m$, but near resonance, the scattering volume is resonantly enhanced: $v_m \approx v^{\rm bg}_m \Delta_m/(B_{0,m}-B)$, where $B_{0,m}$ and $\Delta_m$ are the position and width of the resonance (see Methods and Fig.~\ref{fig:cartoon}). In contrast, $R_m>0$ is only weakly field-dependent~\cite{Ticknor2004,Levinsen:2007gy,Jona2008}.
Since the resonance for $m=z$ is split from the degenerate $m=x$ and $m=y$ resonances, both the strength {\em and} the anisotropy of $p$-wave interactions can be controlled with the magnetic field $\mathbf{B}=B\mathbf{\hat{z}}$~\cite{Regal2003,Ticknor2004,Gunter2005,Peng2014}.

For $v_m>0$, there is a Feshbach dimer state, which is a superposition of $\psi_c$ and $\psi_o$, at energy $E_{\mathrm{d},m}=-\hbar^2 R_m/M v_m$, where $M$ is the atomic mass. For $v_m<0$, the dimer state rises above threshold ($E_{\mathrm{d},m} > 0$) and decays at a rate $\gamma_m= 2 \sqrt{M} R_m E_{\mathrm{d},m}^{3/2} /\hbar^2$ (see Supplementary Information). At low energy, all $p$-wave resonances are narrow ($\hbar \gamma_m \ll E_\mathrm{F}$, where $E_\mathrm{F}$ is the Fermi energy) since $\gamma_m/E_{\mathrm{d},m}$ decreases with energy. In some ways, they resemble narrow $s$-wave resonances~\cite{Hazlett:2012dt,Kohstall:2012kg}, which also have a quasi-bound dimer state above threshold. For the particular $p$-wave resonance we use, dipolar relaxation of $\psi_c$ to more deeply bound states limits the lifetime of the Feshbach dimer to $\lesssim8$\,ms \cite{Ticknor2004,Gaebler2007}.

\smallskip
{\bf Universal $\bm{p}$-wave contact relations}

That $p$-wave interactions might be described with an $\ell=1$ analogue of the contact was conjectured in ref~\onlinecite{Inotani2012}, and recently followed by a full theory in refs~\onlinecite{Yoshida2015,Yu2015}. The most significant structural difference from the $s$-wave contact theory is that each scattering channel has {\em two} contacts, $C_{v,m}$ and $C_{R,m}$, which are the thermodynamic ``forces'' conjugate to $v_m^{-1}$ and $R_m^{-1}$ respectively. We summarize here what can be learned from the contacts, once they are measured or calculated.

{\em A. Thermodynamic identity.} The change in free energy $F$ for a uniform system is
\begin{equation}
dF =  -S \,dT \, + \, \mu \,dN \, - \, P \, dV  
 - \frac{\hbar^2}{2 M} \sum_m C_{v,m} d v_m^{-1} - \frac{\hbar^2}{2 M}  \sum_m C_{R,m} dR_m^{-1},
\label{eq:thermoID}
\end{equation}
where $S$ is entropy, $T$ is temperature, $P$ is pressure, $V$ is volume, $N$ is the total atom number, and $\mu$ is chemical potential. Note that $C_{v,m}$ has units of length, $C_{R,m}$ has units of inverse length, and that both are extensive variables. In a harmonic trap,  $-P \,dV$ is replaced by $2 \omega_\mathrm{osc}^{-1} \langle U \rangle \, d\omega_\mathrm{osc}$, where $\langle U \rangle$ is the potential energy of the cloud and $\omega_\mathrm{osc}$ is the trapping frequency.

{\em B. Correlations.}  At short range, $r \ll k_\mathrm{F}^{-1}$, the many-body wave function has a form that is controlled by two-body physics, but a normalisation that is controlled by the contacts~\cite{Zhang2009}. For instance, the pair correlation function is 
\begin{equation}
g^{(2)}({\bf r}) \to \frac{6 \pi^2}{N} \sum_m Y^2_{1m}(\bm{\hat r}) \left[\frac{C_{v,m} k_\mathrm{F}}{(k_\mathrm{F} r)^4}+\frac{C_{R,m}/k_\mathrm{F}}{(k_\mathrm{F} r)^2}\right] \label{eq:g2}
\end{equation}
in the regime $r_0 \ll r \ll k_\mathrm{F}^{-1}$, where ${\bf r}={\bf r}_1-{\bf r}_2$ is the relative coordinate, $\bm{\hat r}=\bm{r}/r$, and $Y_{1m}$ are the spherical harmonics for $\ell=1$.

{\em C. Momentum distribution.} The contacts also constrain the asymptotic form of the momentum distribution, $n_{\bm k}$. Indeed the $s$-wave contact is often defined to be the high-$k$ limit of $k^4 n_k$~\cite{Tan2008:01,Tan2008:02}. For $p$-waves, in the asymptotic regime $k_\mathrm{F} \ll k \ll 1/r_0$, $n_{\bm k}$ has two components~\cite{Inotani2012,Yoshida2015,Yu2015,Zhou:2016}: \begin{equation} \label{eq:nk}
n_{\bm k} \to \frac{16\pi^2 }{V} \sum_m Y_{1m}^2(\bm{\hat k}) \left[\frac{C_{v,m}}{k^2}+\frac{2 C_{R,m}}{k^4}\right] ,
\end{equation}
where $\bm{\hat k}=\bm{k}/k$ and $V$ is the volume of the system.

{\em D. Spectral response.} The spectral weight of excitations that probe the high-energy or short-range sector of the many-body wave function are also controlled by the contacts~\cite{Chin2005,Pieri2009,Schneider2010,Braaten2010,Stewart2010}. For rf transfer to a non-interacting probe state, the high-frequency tail of the spectral density is
\begin{equation} \label{eq:Irf}
  I(\omega) \to \frac{1}{\pi} \left[C_{v}\sqrt{\frac{M}{\hbar}}\omega^{-1/2} + \frac{3}{2}C_{R}\sqrt{\frac{\hbar}{M}}\omega^{-3/2}\right],
\end{equation}
where $\omega$ is the detuning of the probe frequency from resonance, $C_v \equiv \sum_m C_{v,m}$, and $C_R \equiv \sum_m C_{R,m}$.

{\em E. Fraction of the closed-channel molecules ${f_{\mathrm{c},m}}$.} Close to the Feshbach resonance, where $v_m \gg v_m^{\rm bg}$, $f_{\mathrm{c},m}$ is proportional to $C_{v,m}$:
\begin{equation} \label{eq:ccf}
f_{\mathrm{c},m} = \ell_{\mathrm{c},m}^{-1} C_{v,m}/2 N,
\end{equation}
where $\ell_{\mathrm{c},m} = M \delta\mu\, v_m^{\rm bg} \Delta_m/\hbar^2$~\cite{Yoshida2015}. In this aspect, $C_{v,m}$ is similar to the $s$-wave contact~\cite{Werner2009,Zhang2009}. In contrast, $C_{R,m}$ is an energy-weighted quantity that in the two-channel model also involves atom-dimer interactions (see Supplementary Information).

\begin{figure*}[p]  \includegraphics[width=.9\textwidth]{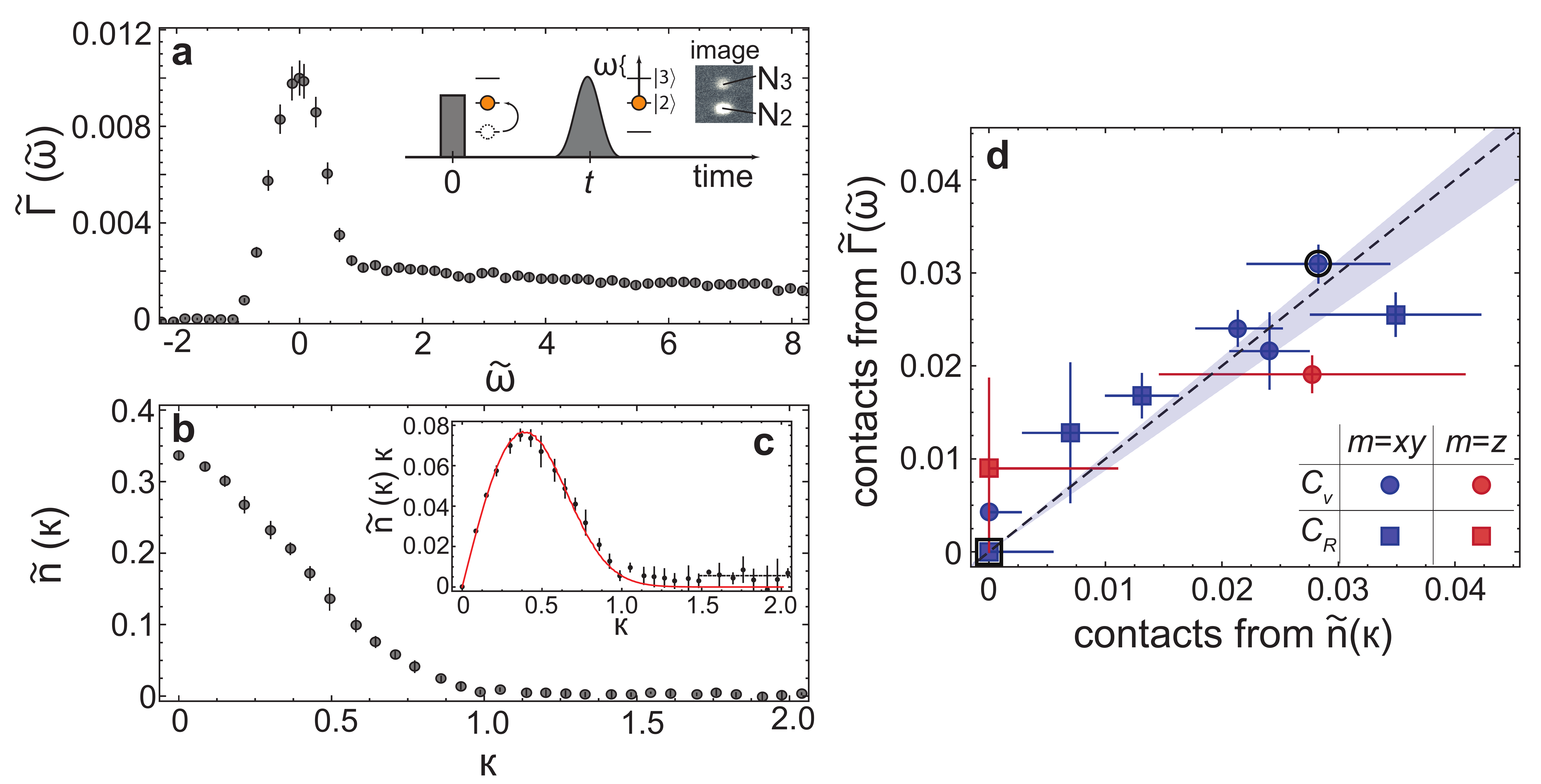}
\caption{{\bf Observation of the $\bm{p}$-wave contacts.} 
{\bf a,} The rf transfer rate $\widetilde \Gamma$ versus frequency $\widetilde \omega$ at $\delta{B}_{xy}=+0.10(2)$\,G and $t=160$\,$\mu$s shows a pronounced tail at $\widetilde\omega > 1$ that can be fit to determine $C_v$ and $C_R$. Each point here corresponds to three repetitions of the time sequence shown as an inset. The transfer fraction is determined from an absorption image after Stern-Gerlach separation of states. {\bf b,c} At the same $\delta{B}_{xy}$ and $t$, the momentum distribution $\widetilde{n}(\kappa)$ also has a visible $\kappa>1$ (i.e. $k>k_\mathrm{F}$) tail, that is not present for a non-interacting gas (red line in c). The dashed line corresponds to the best-fit asymptote, $C_{v,m} k_\mathrm{F}/N = 0.028(6)$ and $C_{R,m}=0.00$. The distribution shown is the average of forty images. {\bf d,} A comparison of contacts determined by $\widetilde \Gamma(\widetilde \omega)$ and by $\widetilde{n}(\kappa)$. Circular markers are $C_{v,m} k_\mathrm{F}/N$ and square markers are $C_{R,m}/k_\mathrm{F} N$; blue markers have $m= xy$ and red markers have $m=z$. Values determined at $\delta{B}_{xy}=+0.10$ are additionally outlined in black. The shaded region shows 1 s.d. uncertainty of the best-fit line, $0.96(7)$, and the dashed line has slope of 1. Error bars in a--d are statistical.}
\label{fig:spectrum} \end{figure*}

\smallskip
{\bf Observation of the $p$-wave contacts}

The primary impediment to the exploration of $p$-wave many-body physics in trapped quantum gases has been atom loss that is faster or comparable to trap-wide equilibration~\cite{Chevy2005,Gaebler2007,Inada2008,Nakasuji2013}. Our experimental approach is to study the gas after a  ``quench'' that quickly initiates enhanced $p$-wave interactions, accomplished with rf pulses. Before each pulse sequence, $^{40}$K atoms are confined in a crossed-beam optical dipole trap, spin-polarised in the lowest hyperfine-Zeeman state $\ket{1}$ and cooled to $T \approx 250$\,nK, which is $\sim 0.2 E_\mathrm{F}/\kB$, above the superfluid critical temperature~\cite{Ohashi2005,Gurarie:2007gs,Inotani2012}. A uniform magnetic field is stabilised at $B=B_{0,m}+\delta B_m$, in the vicinity of a $p$-wave Feshbach resonance for state $\ket{2}$. A resonant 40-$\mu$s $\pi$-pulse transfers all atoms to $\ket{2}$, initiating tuneable $p$-wave interactions. After a variable hold time $t$, the gas is characterised either with rf spectroscopy or with time-of-flight (TOF) imaging, allowing contacts to be measured through relations (\ref{eq:Irf}) or (\ref{eq:nk}) respectively, as shown in Fig.~\ref{fig:spectrum}. Losses restrict $t$ to be short compared to thermalisation times of low-energy or long-wavelength degrees of freedom. However, we find that spectra reach a quasi-steady-state, which likely reflects a local equilibrium.

Radio-frequency spectroscopy probes the gas by transferring a fraction of atoms in $\ket{2}$ to the third-lowest energy state $\ket{3}$, which (like $\ket{1}$) does not have resonantly enhanced interactions. The fractional transfer to $\ket{3}$, $N_3/(N_2+N_3)\equiv N_3/N$, is measured by state-selective absorption imaging, after a magnetic field jump that dissociates any Feshbach dimers. Figure~\ref{fig:spectrum}a shows an rf spectrum taken at $\delta{B}_{xy}=+0.10(2)$\,G. The transfer to $\ket{3}$ is given as a rescaled rate $\widetilde \Gamma(\widetilde \omega) = (E_{\rm{F}}/\hbar) (\pi \Omega^2 t_{\rm{rf}})^{-1} (N_3/N)$, where $\widetilde\omega = \hbar\omega/E_{\rm F}$ is the probe frequency rescaled by $E_{\rm F}$, $\hbar\Omega/2$ is the transition matrix element and $\Omega t_{\rm{rf}}$ is the pulse area. The latter is chosen to be small enough to probe the transition in the linear regime for $\widetilde\omega \gg 1$, where $\widetilde\Gamma(\widetilde\omega) \propto I(\widetilde \omega)$. The high-frequency tail fits well to equation~(\ref{eq:Irf}), and is used to determine $C_v$ and $C_R$ (see Methods and SI for details).

The momentum distribution is measured by resonant absorption imaging of the cloud after a $5.5$\,ms time-of-flight expansion. For these measurements, no rf pulse is applied at $t$; instead, the field is rapidly jumped away from the $p$-wave resonance, preserving the interacting momentum distribution, which determines the ballistic flight after release from the trap. Figure~\ref{fig:spectrum}b shows the normalized distribution $\widetilde n(\kappa)$ observed at $\delta B_{xy} = +0.10(2)\,{\rm G}$ versus $\kappa = \sqrt{(k_x^2+k_y^2)}/k_{\rm F}$, after azimuthal averaging in the image plane. Inherent to imaging is also a line-of-sight integration of $n(\vec{k})$, so that the high-momentum scaling of $\widetilde{n}(\kappa)$ is $\propto \kappa^{-1}$ for the $C_{v,m}$ term and $\propto \kappa^{-3}$ for the $C_{R,m}$ term (Methods). Figure~\ref{fig:spectrum}c shows that the leading order appears as an asymptotic plateau in $\kappa \times \widetilde{n}(\kappa)$. A full fit with both terms is used to determine $C_v$ and $C_R$.

Figure~\ref{fig:spectrum}d compares the $p$-wave contacts determined from $\widetilde{\Gamma}(\widetilde\omega)$ and $\widetilde{n}(\kappa)$, across a range of magnetic field values. The dimensionless $C_v k_\mathrm{F}/N$ and $C_R/ k_\mathrm{F} N$
are scaled by $k_\mathrm{F}$ calculated using the peak density of a non-interacting gas, but 
results should be understood as an average over an inhomogeneous trapped ensemble~\cite{Sagi2012}. Since the contacts are only revealed in the asymptotic part of the distribution, analysis involves a low-energy cutoff, the systematic effect of which is studied in SI. Our analysis also assumes both $C_v$ and $C_R$ are nonnegative, but the possibility of $C_R<0$ is also discussed in the SI.

The correlation between the two observables is $0.96(7)$, as determined by the slope of a best-fit line with no offset. This agreement, in addition to the observation of the predicted asymptotic scaling of equations~(\ref{eq:nk}) and (\ref{eq:Irf}), is strong evidence that the $p$-wave contact relations are valid.

\begin{figure*}[p]  \includegraphics[width=0.9\textwidth]{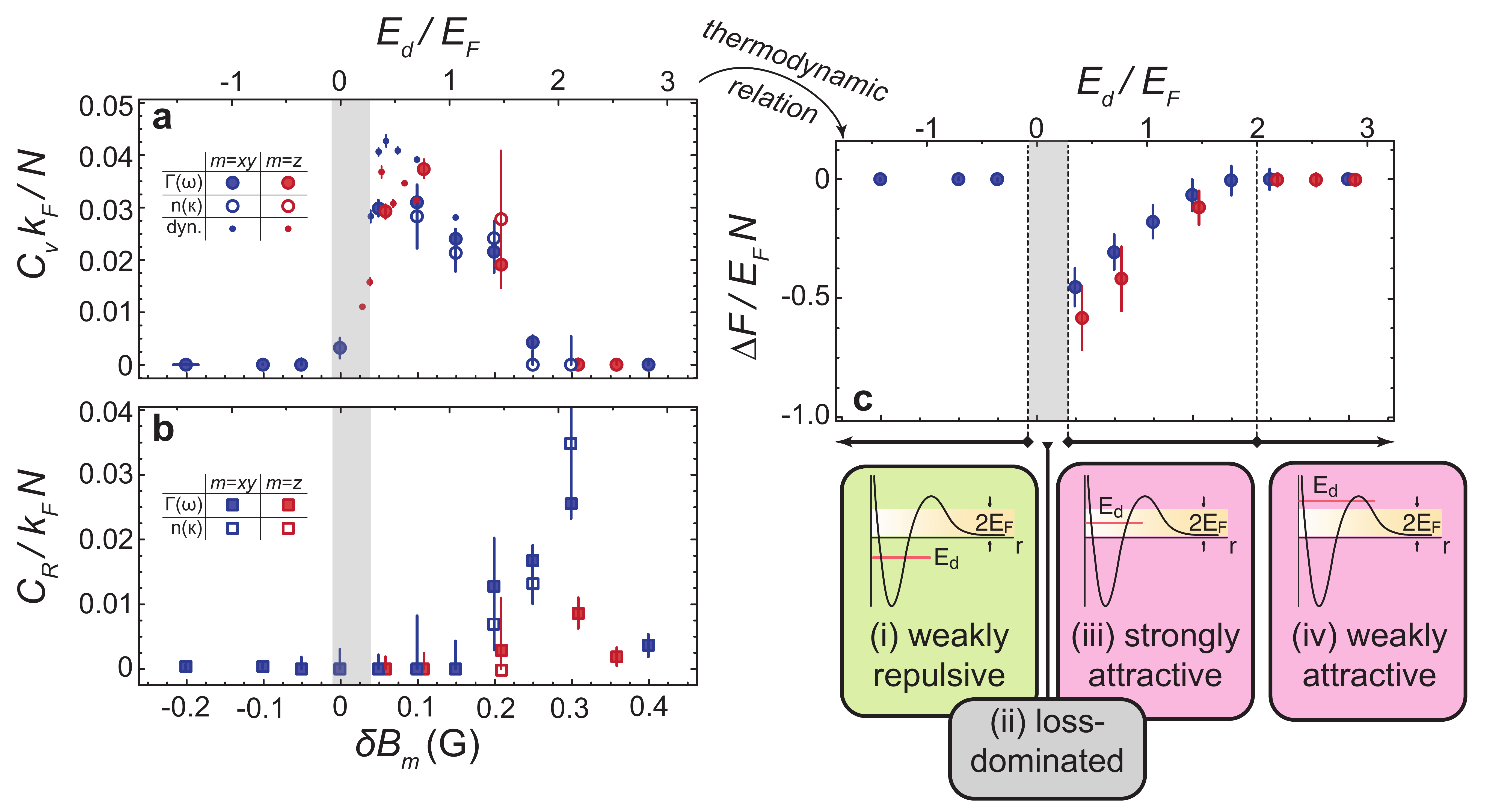}
\caption{{\bf The $\bf{p}$-wave contacts near two Feshbach resonances.} {\bf a,} $C_v k_\mathrm{F}/N$ and {\bf b,} $C_R/k_\mathrm{F} N$, versus magnetic field $\delta B_m$ (lower axes) or dimer energy $E_{\mathrm{d},m}/E_\mathrm{F}$ (upper axes). Data are shown from both the $m=xy$ (blue) and $m=z$ (red) resonances, from rf spectra (filled) and momentum distributions (open). Momentum spectroscopy is shown for a smaller range of $\delta B_m$, due to limited signal-to-noise. Most data is taken at $t=160$\,$\mu$s, however long-time asymptotes from fits to dynamical data (see Fig.~\ref{fig:dynamics}) are also shown as smaller filled points. {\bf c,} Numerical integration of measured $C_v$ gives the shift of free energy $\Delta F = F - F_\mathrm{bg}$ due to near-resonant interactions (see Methods for the details of the numerical integration). Data is referenced to $F_\mathrm{bg} = F(B_\mathrm{max})$ for $\delta B_m>0$; and to $F_\mathrm{bg} = F(B_\mathrm{min})$ for $\delta B_m<0$. Illustrations depict the dimer energy, compared to the range of collision energies available in the Fermi sea. Error bars are statistical; see text for a discussion of systematic uncertainty.}
\label{fig:contacts}
\end{figure*}

\smallskip
{\bf Field dependence of the $p$-wave contacts}

Figures~\ref{fig:contacts}a,b show the $p$-wave contacts versus $\delta B_m$ near both the $m=xy$ and the $m=z$ resonances. The data includes contacts determined at $t=160\us$ from $\widetilde\Gamma$, at $t=160\us$ from $\widetilde n$, and asymptotic values from $\widetilde\Gamma$ versus $t$. The variable-t data (discussed in more detail below) also identifies a loss-dominated regime for $0.00(2) \,\mathrm{G} \leq \delta B_m \leq 0.04(2)$\,G, outside of which contacts reach a steady-state value despite atom loss of up to 20\%.

We observe a pronounced asymmetry about each Feshbach resonance: significant contacts are only observed for $\delta B_m >0$. $C_v$ is largest close to resonance, decreases with $\delta B_m$, and vanishes beyond $\delta B_m \approx 0.3$\,G, where $E_{\mathrm{d},m}/E_{\rm F} \approx 2$. $C_R$ instead peaks at $\delta B_m \approx 0.3$\,G  before abruptly falling to zero for larger fields. 

Some of these salient features can be explained by a simple model, in which $N_\mathrm{d} = \sum_m N_{d,m}$ non-interacting closed-channel dimers are in equilibrium with $N_{\rm f}$ free fermions. Each dimer has $C_{v,m}=2 R_m$ and $C_{R,m}=-2 R_m^2/ v_m$, but free fermions make no contribution to the contacts. Since the $m=xy$ and $m=z$ resonances are well separated, $C_v k_\mathrm{F}/N \approx 2 k_\mathrm{F} R_m (N_\mathrm{d}/N)$ and $C_R/(k_\mathrm{F} N) \approx 2 k_\mathrm{F} R_m (E_{\mathrm{d}} /E_\mathrm{F}) (N_{\mathrm{d}}/N)$. The assumption of equilibrium gives $N_\mathrm{d}=(N/2)(1- (E_{\mathrm{d},m}/2 E_\mathrm{F})^3)$ in a harmonic trap at zero temperature~\cite{Gurarie:2007gs}. This model would predict that both $C_v$ and $C_R$ are the same near the $xy$ and the $z$ resonances, that $C_v\to0$ and $C_R\to0$ as $E_{\mathrm{d},m} \to 2E_{\rm F}$, and that a fully dimerised gas would have $C_v k_\mathrm{F} / N \approx 0.04$, since $k_\mathrm{F} R_m \approx 0.04$ in typical conditions. The additional factor of $(E_{\mathrm{d},m} /E_\mathrm{F})$ in $C_R$ gives $C_R =0$ at resonance and at peak $C_R/(k_{\rm{F}} N)\approx0.06$ at $E_{\mathrm{d},m}/ E_{\mathrm{F}} \approx 1.6$.

Although this model does explain the peak value of $C_v$ and the range of $\delta B_m$ at which significant contacts are seen, it does not explain the value or location of the maximum in $C_R$. A more realistic model would include finite temperature, and interactions between dimers, between atoms, and/or between atoms and dimers. For instance, resonant enhancement of atom-dimer interactions have been seen in a three-body calculation~\cite{Levinsen:2007gy,Jona2008}.

Independent of any particular microscopic model, but assuming adiabaticity, we can understand the thermodynamic implications of the observed contacts using equation~(\ref{eq:thermoID}). The change in free energy $F$ versus $\delta B_m$ is given by the integral of $C_v$ over $v_m^{-1}$, assuming all other variables are constant. The contribution of $C_R$ is not significant (Methods). The inferred $\Delta F$ is shown in Fig.~\ref{fig:contacts}c. The values shown have several possible systematic errors. First, some of the other variables that determine $F$ are varied by $\delta B_{m}$: $N$ decreases due to loss, and $T$ increases by $\sim 0.05\,E_\mathrm{F}/\kB$ near resonance. A second and more significant error may lie in the calibration of number and rf power, which combine to give a 30\% systematic uncertainty in $\Delta F \equiv F - F_{\rm bg}$. Finally, equilibration is likely to be only local, and not trap-wide. Despite these uncertainties, the integrated data is sufficient to demonstrate several qualitative regimes:

(i) Below resonance ($E_\mathrm{d}<0$), the gas is weakly repulsive, with $0 \leq \Delta F \ll E_\mathrm{F}$. Here, resonant scattering is inaccessible to free particles, and the gas remains on the ``upper branch''~\cite{Pricoupenko2006,Shenoy:2011dm}. Few or no dimers are formed, because energy-conserving two-body collisions cannot produce a dimer with a finite binding energy. Instead, the gas has weakly repulsive $p$-wave interactions.

(ii) At resonance, we do not extract a value for $F$, because a steady-state in $C_v k_\mathrm{F}/N$ is not achieved, as discussed in the next section. However, the discontinuity in $F$ between regime (i) and regime (iii) implies that the systems shifts from upper to lower branch in this region.

(iii) Above resonance, in the range $0.25 \lesssim E_\mathrm{d}/E_\mathrm{F} \lesssim 2$, we infer a reduction in $F$ per particle approaching half the Fermi energy near resonance. The significant reduction of free energy is partially explained by the formation of dimers, whose binding energy could contribute up to $\Delta F = - 3 E_\mathrm{F}/4$ in a harmonic trap~\cite{Gurarie:2007gs}. 
Additional contributions to $\Delta F$ include dimer-dimer or atom-dimer interactions. Accompanying the large $\Delta F$ in this regime are the largest observed contacts, and therefore the strongest $p$-wave correlations, as described be equation (\ref{eq:g2}).

(iv) Farther above resonance, the scattering resonance at $E_\mathrm{d}$ exceeds the maximum collision energy $2 E_\mathrm{F}$ in a zero-temperature Fermi sea, leaving primarily non-resonant interactions between atoms. In this regime, $p$-wave interactions are weakly attractive: 
$0 \leq -\Delta F \ll E_\mathrm{F}$.

\begin{figure*}[p]  
\includegraphics[width=.95\textwidth]{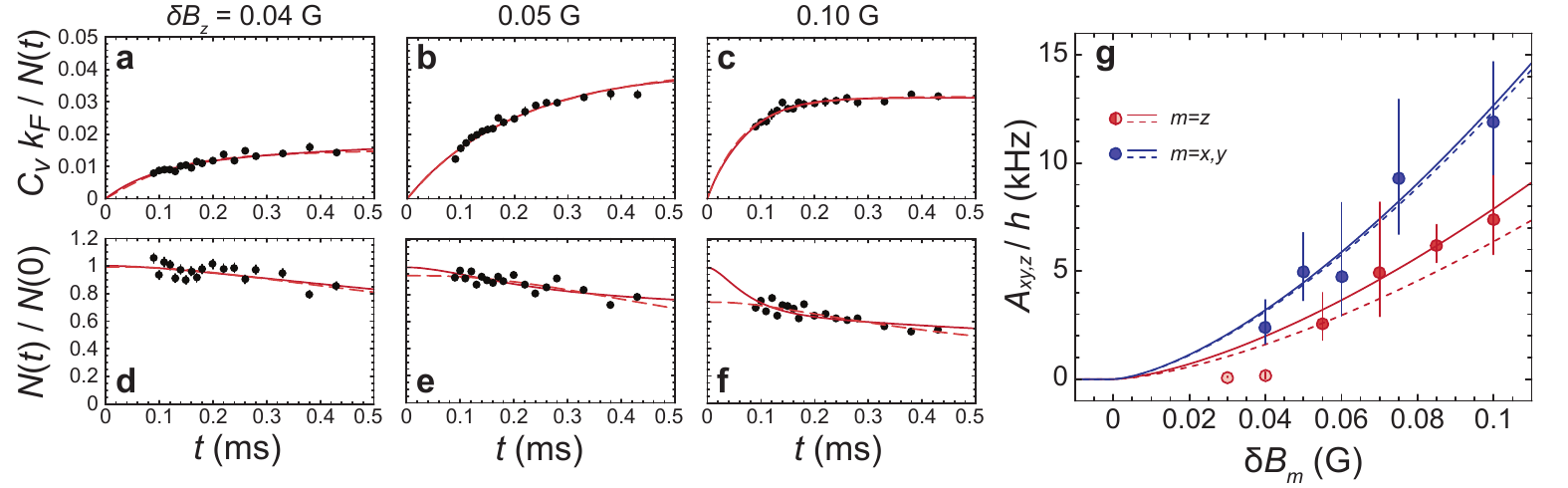}
\caption{{\bf Dynamics of the $p$-wave contact and atom number close to resonance.} $C_v k_\mathrm{F}/N$ ({\bf a-c}) and $N(t)/N(0)$ ({\bf d-f}) are shown for $B = 198.83\,{\rm G}$ ({\bf a, d}), $198.845\,{\rm G}$ ({\bf b, e}), and $198.89\,{\rm G}$ ({\bf c, f}). Error bars in {\bf a-f} represent the standard deviation over six separate measurements. Solid lines represent best fits to equation~(\ref{eq:rateeq}); dashed lines consider instead a dimer-dimer loss term discussed in the Supplementary Information. $N(t)$ is normalised to the extrapolated $N(0)$. We vary $A_m$, $L_{\rm fd}$, and $C_v/N_{\rm d}$ to fit the solution of equation~(\ref{eq:rateeq}) to our data and iteratively determine a self-consistent $B_{0,m}$ for either resonance (see Supplementary Information). We find $B_{0,z}=198.792(13)\,{\rm G}$ and $B_{0,xy}=198.301(11)\,{\rm G}$, which are used to determine the $\delta B_m$ values reported in this work. {\bf g,} The dimer association rates $A_{xy}$ (blue circles) and $A_z$ (red circles) versus magnetic field. The open circles indicate a loss-dominated solution. Solid lines in {\bf g} are fits to $A_{xy}$ and $A_z$ with assumed proportionality to $\gamma_m$, finding $A_{xy}/\gamma_{xy} = 16.3(3.7)$ and $A_z/\gamma_z = 9.9(2.5)$. Dotted lines represent the theoretical prediction of $16\gamma_{xy}$ (red) and $8\gamma_z$ (blue), respectively. Error bars in {\bf g} correspond to $\sigma$ confidence intervals.}
\label{fig:dynamics} 
\end{figure*}

\smallskip
{\bf Dynamics of $\bf{C_v}$ after the quench}

Evolution of the many-body wave function is required to adjust to the $p$-wave interactions initiated by transferring atoms from $\ket{1}$ into $\ket{2}$. We are able to observe these dynamics in the range $0.02\,{\rm G} \leq \delta B_m \leq 0.10\,{\rm G}$, by varying the time $t$ at which we measure the contacts. For an isolated pair of atoms, the lifetime of the quasi-bound state $\gamma_m^{-1}$ would set the key time scale (between 0.2\,ms and 1\,ms here). However, we find that the contact develops at a rate twenty to thirty times larger than $\gamma_m$.

From Fig.~\ref{fig:contacts} we know that only $C_v$ contributes to $\widetilde{\Gamma}$ for $\delta B_m \lesssim 0.1\,{\rm G}$. Measurement of $\widetilde{\Gamma}$ at a single rf frequency (typically $\widetilde{\omega} \sim 5$) is used to determine $C_v k_\mathrm{F}/N$. We also record the atom number $N$ at each $t$. Figures~\ref{fig:dynamics}a--f show the time-resolved measurements of $C_v k_\mathrm{F}/N$ and $N$ at various $\delta B_z$. We observe that $C_v k_\mathrm{F}/N$ rises to an apparent steady-state value, while $N$ decays relatively slowly. The initial growth rate of $C_v k_\mathrm{F}/N$ increases with $\delta B_m$, until it can no longer be resolved, at $\delta B_m \gtrsim +0.10\,{\rm G}$.

Near resonance, the wave function of each Feshbach dimer is dominated by $\psi_c$~\cite{Fuchs2008,Gubbels2007}, allowing us to interpret the contact dynamics through a multi-channel model and equation~(\ref{eq:ccf}). Three closed channels of $p$-wave dimers are initially empty,  but come to equilibrium with the initially populated open channel of atoms, but all contribute to the observed atom number $N = N_{\rm f} + 2\sum_m N_{{\rm d},m}$. We calculate the time evolution of $N_{\rm f}$ and $N_{{\rm d},m}$ with rate equations that omit coherence between the channels but include collisional loss:
\begin{align}
 &\dot{N}_\mathrm{f}  = -\sum_m \Big(2A_m N_{\rm f} -2\gamma_m N_{{\rm d},m}+ L_{{\rm fd},m}\,N_{\rm f} N_{{\rm d},m} \Big), \nonumber \\
 &\dot{N}_{{\rm d},m}  = A_m N_{\rm f} - \gamma_m N_{{\rm d},m} - L_{{\rm fd},m}\,N_{\rm f} N_{{\rm d},m}, \label{eq:rateeq} 
\end{align}
where $A_m$ is the dimer association rate, and $L_{{\rm fd},m}$ is the fermion-dimer loss coefficient. We constrain the model to have $N_{{\rm d},x}=N_{{\rm d},y}=0$ near the $z$ resonance, and $N_{{\rm d},z}=0$ near the $xy$ resonance. We furthermore combine the degenerate modes at the $xy$ resonance to $N_{{\rm d},xy} \equiv N_{{\rm d},x} + N_{{\rm d},y}$, $A_{xy} \equiv A_x + A_y$, and $\gamma_{xy} \equiv \gamma_{x,y}$. We fit the solution of equation~(\ref{eq:rateeq}) to the data, finding good agreement (see Figs.~\ref{fig:dynamics}a--f). The rapid growth of $N_{{\rm d},m}$ for $\delta B_{m} > 0$ was observed directly in refs~\onlinecite{Jin2008,Inada2008}.

With $L_{{\rm fd},m}=0$, equation~(\ref{eq:rateeq}) would lead to a dimer population that asymptotically tends towards $N_{{\rm d},xy} = (A_{xy}/\gamma_{xy})N_{\rm f}$ and $N_{{\rm d},z} = (A_{z}/\gamma_{z})N_{\rm f}$. The associated equilibration rates are 
\begin{equation}
1/\tau_{xy,z} = \gamma_{xy,z} + 2A_{xy,z}. 
\end{equation}
With loss present, we can distinguish between regimes (ii) and (iii). If $\tau_m\,L_{{\rm fd},m} \ll 1$, the system reaches a quasi-steady state from which it decays slowly. This scenario is realised in Fig.~\ref{fig:dynamics}b,e and Fig.~\ref{fig:dynamics}c,f, and corresponds to regime (iii) in Fig.~\ref{fig:contacts}c. If, however, $\tau_m \,L_{{\rm fd},m} \gg 1$, the loss rate of the system is faster than the association rate, and equilibration between $N_{\rm d}$ and $N_{\rm f}$ is inhibited. As a consequence, $N_{\rm d}$ reaches a significantly reduced steady-state value on a time scale that is dominated by the loss rate. In turn, atom loss occurs on the time-scale of the dimer association. This case, corresponding to regime (ii) in Fig.~\ref{fig:contacts}c, occurs closer to resonance (see Fig.~\ref{fig:dynamics}a,d and data at $\delta B_z = +0.04$\,G and $\delta B_z = +0.03$\,G in Fig.~\ref{fig:dynamics}g), because $1/\tau_{xy,z} \to 0$ for $\delta B_m \to 0$. The SI provides further details of how these two regimes are distinguished, and demonstrates robustness against details of the loss model. 

Figure \ref{fig:contacts}a includes the asymptotic values of $C_v$ (smaller points). They are within uncertainty of $C_v$ at $t=160\us$ for $\delta B_m \geq 0.1$\,G, but provide an upwards correction for $\delta B_m \sim 0.05$\,G. With the inclusion of this correction, $C_v$ decreases monotonically with $\delta B_m$ throughout regime (iii).

Using only data from regime (iii), we find that $A_m$ are proportional to $\gamma_m$, with best-fit ratios $A_{xy}/2 \gamma_{xy} = 8.1(1.8)$ and $A_z/\gamma_z = 9.9(2.5)$ (see lines in Fig.~\ref{fig:dynamics}g). These ratios are consistent with a perturbative treatment of resonant closed-channel molecular formation in a $T=0$ Fermi cloud that predicts $A_m \to 8 \gamma_m$ as $1/v \to 0^-$ (SI). The consistency between the dynamical response of $C_v$ and a model of dimer population supports the validity of equation (\ref{eq:ccf}).

\smallskip
{\bf Conclusions}

While studied here for a Fermi gas in a metallic state, the $p$-wave contacts are expected to be  ``universal'' in that they hold for any type of particle, whether boson or fermion, in any dimensionality, and in any state (superfluid or normal), so long as interactions are short-range and $p$-wave. Further tests of universality should include comparison to direct measurements of energy, structure factor, and dimer number, as has been done for the $s$-wave contact~\cite{Partridge2005,Stewart2010,Kuhnle2010,Navon:2010ix,Kuhnle2011}.

We have also identified a regime where the atom-dimer equilibration is much faster than loss. After equilibration, strong $p$-wave correlations persist for at least half a millisecond, eventually limited by dipolar and three-body loss rates. Searching for the onset of pair condensation in this dynamical window at a low temperature would be of great interest, especially in two-dimensional systems, where some loss mechanisms are less significant~\cite{Levinsen2008}, and the possibility of chiral order exists~\cite{Read:2000iq,Levinsen:2007gy}.

\clearpage
\singlespacing

{\bf Methods}

{\em Sample preparation}.
Spin-polarised $^{40}$K atoms are cooled sympathetically with bosonic $^{87}$Rb atoms. The $\ket{1},\,\ket{2}$, and $\ket{3}$ states of $^{40}$K refer to the high-field states adiabatically connected to the low-field $m_f=-9/2,\,-7/2$, and $-5/2$ states of the $f=9/2$ hyperfine manifold of the electronic ground state, where $f$ and $m_f$ denote the total angular momentum and the corresponding magnetic quantum number respectively. At the end of cooling, residual $^{87}$Rb atoms are removed with a resonant light pulse leaving only $^{40}$K atoms in state $\ket{1}$, held in a cross-beam optical dipole trap. For spectroscopic data, $N = 3.6(4)\times 10^4$, while dynamics data was taken with $N = 2.5(5)\times 10^4$. The $\pi$ pulse initialising dynamics is $\approx 98\%$ efficient, so that we assume $N_2 = N_1$. The mean oscillation frequency in the trap is $\bar{\omega}_\mathrm{osc}/2\pi=440$\,Hz.

The ideal-gas Fermi energy in a harmonic trap is $E_\mathrm{F} = \hbar \bar{\omega}_\mathrm{osc} (6 N)^{1/3}$, here $\sim h \times 30\,$kHz, and is also the local Fermi energy at the centre of the trap. Thermometry is based on fitting an ideal-gas Fermi distribution to an absorption image taken at high magnetic field after release from the trap. To measure the temperature after a $160\,\mu$s hold time in $\ket{2}$, we jump the field rapidly to 209\,G, turn off the trap, and transfer all atoms to $\ket{1}$ during time-of-flight. At $\delta B_{xy}=0.20$\,G, the apparent temperature rise is $\approx 50$\,nK. Accompanied with a 25\% number loss, this is an increase in reduced temperature $\kB T/E_\mathrm{F}$ of $\sim$0.05. We note that with such short hold times, the gas is likely out of thermal equilibrium.

State-resolved absorption images are taken at high magnetic field, after Stern-Gerlach separation of states $\ket{1}$, $\ket{2}$, and $\ket{3}$. Subsequent state transfer in a residual gradient field is used to map any population of interest to $\ket{1}$, since neither of the other states have an accessible cycling transition. For spectroscopic measurements, we optimise and calibrate this sequence to determine $N_2$ and $N_3$.

The magnetic field values are calibrated with the frequency of the $\ket{1}$--$\ket{2}$ transition, measured spectroscopically with a 120-$\us$ pulse to an accuracy of $\sim 1$\,kHz. The short-term accuracy of field values is $\sim 2$\,mG. Over one spectroscopic or dynamic data series, the magnetic field drifts by less than $10$\,mG.

\smallskip
{\em Analysis of rf spectra}.
The high-frequency tail in the spectral density $I(\omega)$ of the rf transition to a non-interacting probe state ($\int\!I(\omega)\,d\omega = N$) is given by equation~(\ref{eq:Irf}). We measure the transfer rate $\Gamma(\omega) = N_{3}(\omega)/t_{\rm rf}$, with $t_{\rm rf}$ the length of the spectroscopy pulse. The transfer rate is proportional to the spectral density and obeys the sum rule $\int\!\Gamma(\omega)\,d\omega = \Omega^2 \pi N/2$. We therefore identify $\Gamma(\omega) = \Omega^2\pi I(\omega)/2$. The scaled frequency $\widetilde{\omega}$ and normalised transfer rate $\widetilde\Gamma$ as defined in the main text are normalised to $\int\!\widetilde\Gamma(\widetilde{\omega})\,d\widetilde{\omega} = 1/2$. 

With these definitions, the high-frequency tail of the normalised transfer rate is
\begin{equation} \label{eq:transfer}
\widetilde{\Gamma} \rightarrow\frac{1}{2^{3/2}\pi} \frac{C_v k_\mathrm{F}}{N}\,{\widetilde\omega}^{-1/2} 
+ \frac{3}{2^{3/2} \pi} \, \frac{C_R}{k_\mathrm{F} N}\,{\widetilde\omega}^{-3/2}.
\end{equation}
Close to the $p$-wave resonances where $C_R/k_\mathrm{F}^2 \ll C_v$, we measure  $C_v$ using $N_3(\omega)/N$ at a single frequency value $\omega$:
\begin{equation} \label{eq:ssSM}
  \frac{C_v k_\mathrm{F}}{N} \simeq \frac{2}{\Omega^2 t_{\rm rf}}\,\sqrt{\frac{\hbar}{M}}\,k_\mathrm{F}\,\sqrt{\omega}\,\frac{N_3(\omega)}{N}\,.
\end{equation}
This single-frequency measurement technique is used in the data presented in Fig.~\ref{fig:dynamics}.

We calibrate the strength of rf fields by driving Rabi oscillations between $\ket{1}$ and $\ket{2}$ at 209\,G. At the peak power used in this experiment, we find a Rabi frequency of approximately 70\,kHz with a typical uncertainly of 15\%.

The spectroscopic pulse has a Blackman envelope to minimise the frequency sidebands in the spectra. For a given pulse length $t_\mathrm{rf}$, the Blackman pulse area is a factor of $\approx$0.4266 different from that of a square pulse with the same length and peak power. The effective Rabi frequency that is used to normalise our rf spectra is $\Omega/2 \pi=30(5)$\,kHz. Within the measurement window, we find $\int\!\widetilde\Gamma(\widetilde{\omega})\,d\widetilde{\omega} \simeq 0.41\,(0.44)$ for $\delta B_{xy}=+$0.05\,G\,(-0.10\,G), which agrees with the sum rule to within systematic error. Note that when we take normalisable spectra the Rabi frequency must be reduced for $|\widetilde\omega| \lesssim1$, to keep the coupled fraction small.

\smallskip
{\em Analysis of momentum distributions}.
The momentum distribution in Fig.~\ref{fig:spectrum}b,c is obtained from a time-of-flight  absorption image. The imaging beam propagates along the $z$ direction, parallel to the Feshbach field. The optical density of the cloud is proportional to the column-integrated density $\rho(x,y,t_{\rm{TOF}})=\int dz\rho({\bf r},t_{\rm{TOF}})$. For long time of flight, the initial cloud size is rendered unimportant and $\rho({\bf r},t_{\rm{TOF}})\propto n({\bf k})$. We normalize $\rho$ by its pixel sum to obtain $\widetilde{n}({\bf k})$ with $\int\widetilde{n}({\bf k})d^3k=1$. Signal-to-noise is further improved by azimuthal averaging, yielding a distribution $\widetilde{n}(\kappa)$ , where $\kappa \equiv \sqrt{k_x^2 + k_y^2}/k_\mathrm{F}$. Thirty to forty images are taken, and averaged. The contacts are determined with a fit to 
\begin{eqnarray} \label{eq:momdistscaled}
\widetilde{n}(\kappa) \to \frac{3}{4\pi}\frac{C_{v}k_\mathrm{F}}{N}\kappa^{-1}+\frac{3}{8\pi}\frac{C_{R,z}+3C_{R,xy}}{Nk_\mathrm{F}}\kappa^{-3}
\end{eqnarray}
While the coefficient of the leading order is proportional to $\sum_m C_{v,m}$, just as in rf spectra, the sub-leading order is not simply proportional to $\sum_m C_{R,m}$. We interpret the data assuming that $C_{R,z}=0$ near the $xy$ resonance, and that $C_{R,xy}=0$ near the $z$ resonance.

\smallskip
{\em $p$-wave scattering parameters}.
For the $p$-wave Feshbach resonance in state $\ket{2}$ near 198.5\,G we parameterise $v_m$ and $R_m$ as a function of magnetic field magnitude $B$ using
\begin{align}
v_m & = v^{\rm bg}_m\left(1-\frac{\Delta_m}{\delta{B}_m}\right) \quad \mbox{or} \label{eq:scattvol} \\
\frac{1}{v_m } & \approx \frac{1}{v^{\rm bg}_m } \left[ -\frac{\delta{B}_m}{\Delta_m} + \left( \frac{\delta{B}_m}{\Delta_m} \right)^2 + \mathcal{O} \left( \frac{\delta{B}_m}{\Delta_m} \right)^3 \right] \nonumber \\
\frac{1}{R_m} &= \frac{1}{R_m^{\rm bg} } \left[1 + \frac{\delta{B}_m}{\Delta_{R,m}} + 
\mathcal{O} \left( \frac{\delta{B}_m}{\Delta_{R,m}} \right)^2 \right] \label{eq:scattrange}
\end{align}
where $\delta{B}_m = B- B_{0,m}$. From ref \onlinecite{Ticknor2004}, we use the values 
$v^{\rm bg}_z =  (101.6\,a_0)^3$,
$v^{\rm bg}_{xy} = (96.74\,a_0)^3$,
$\Delta_z =  21.95 {\rm G} $,
$\Delta_{xy} =  24.99\,{\rm G}$,
$R^{\rm bg}_z =  47.19\,a_0$,
$R^{\rm bg}_{xy} =  46.22\,a_0$,
$\Delta_{R,z} =  -18.71\,{\rm G}$, and
$\Delta_{R,xy} =  -22.46\,{\rm G}$. From our own measurements, we use
$B_{0,z} =198.792(13){\rm G}$ and
$B_{0,xy} =  198.301(11){\rm G}$. Equations~(\ref{eq:scattvol}) and (\ref{eq:scattrange}) are matched to ref~\onlinecite{Ticknor2004}, neglecting the next order in $\delta B_m / \Delta$. However, since $|\delta B_m| < 0.4$\,G in this work, the next-order term has a relative magnitude $10^{-6}$ for $v$ and $10^{-3}$ for $R$.  The binding energies measured in ref~\onlinecite{Gaebler2007} matches the calculated $\Delta v_{\rm bg}/R^{\rm bg}$ to 3\%, but further measurements are needed to constrain, for instance, $\Delta_R$. Our measurements of $B_{0,z}$ and $B_{0,xy}$ are consistent, within error, to those found in ref~\onlinecite{Gaebler2007}. Both results may share a systematic offset due to the differential polarisability of the open and closed channel.

\smallskip
{\em Numerical integration to find $\Delta F$}. As a function of magnetic field, the reversible change in free energy is
\begin{equation}
\frac{d F}{d B} = \sum_m \frac{\partial F}{\partial ({1}/{v_m})} \frac{d ({1}/{v_m})}{d B}
+ \frac{\partial F}{\partial (1/R_m)} \frac{d (1/R_m)}{d B} \nonumber
\end{equation}
Assuming the $xy$ resonance is well isolated from the $z$ resonance, all terms in this equation can be evaluated with equation~(\ref{eq:thermoID}), measurements of $C_{v,m}$ and $C_{R,m}$, and the parameterisations (\ref{eq:scattvol}) and (\ref{eq:scattrange}). The dominant contribution comes from the scattering volume:
\begin{equation} \label{eq:dFdBv}
\frac{d F}{d B} \approx \frac{\hbar^2 C_v}{2 M v^{\rm bg}_m \Delta_m} 
\left[ 1 - 2 \frac{\delta{B}_m}{\Delta_m} + \mathcal{O} \big( \frac{\delta{B}_m}{\Delta_m} \big)^3 \right]. 
\end{equation}

The change in energy between two magnetic field values, $B_1$ and $B_2$, is given by $F(B_2) - F(B_1) = \int_{B_1}^{B_2} \frac{dF}{dB} dB$. To evaluate the shift in  energy due to near-resonant interactions, we calculate the numerical integrals
\begin{align}
F(B) \approx 
& F(B_\mathrm{min}) + \int_{B_\mathrm{min}}^{B} \! \frac{\hbar^2 C_v(B') }{2 M v^{\rm bg} \Delta}  dB' \quad \mbox{for $B < B_0$} \\
F(B) \approx 
& F(B_\mathrm{max}) - \int_{B}^{B_\mathrm{max}} \frac{\hbar^2 C_v(B') }{2 M v^{\rm bg} \Delta} dB' \quad \mbox{for $B > B_0$} 
\end{align}
where we have suppressed the ${m}$ subscripts for clarity, but are assuming that the $z$ and $xy$ resonances are isolated. To estimate the statistical uncertainty, we repeat the integration with data normally distributed around the mean values of $C_{v}$ at each $B$. The reported error is the standard deviation of the resultant $F(B)$. The next order in $\delta B_m / \Delta$ given in equation~(\ref{eq:dFdBv}) can easily be included, but $|\delta B_m / \Delta|$ is $\leq 0.01$ when $C_v$ is above the noise floor of spectroscopy, so the resultant shift is not significant.

The $B$-dependence of $F$ due to the change in $R$ is 
\begin{equation} \label{eq:dFdBR}
\frac{d F}{d B}\bigg|_{T,v} \approx \frac{-\hbar^2 C_R}{2 M R_m^{\rm bg} \Delta_{R,m}}
\end{equation}
Across the range in which we observe $p$-wave contacts, this contribution to $\Delta F$ is small. One can understand this by comparing equations~(\ref{eq:dFdBv}) to (\ref{eq:dFdBR}): since $|\Delta_R| \sim |\Delta|$, even if $C_R / k_\mathrm{F} N$ is comparable to $C_v k_\mathrm{F}/N$, the ratio of contributions to $dF/dB$ near resonance is roughly $k_\mathrm{F}^2 v^{\rm bg}/R^{\rm bg}$, which is $\sim 10^{-2}$ for typical experimental parameters.

{\bf Acknowledgements} We thank F.~Chevy for discussion and shared notes concerning $F$ versus $\delta B_m$. We also thank N.~Zuber for experimental assistance, and J.~Bohn, B.~Ruzic, Shina Tan, Edward Taylor, Pengfei Zhang, and Qi~Zhou for discussion. This work was supported by AFOSR under FA9550-13-1-0063, ARO under W911NF-15-0603, the Croucher Foundation, RGC under 17306414, NKBRSFC, NSERC, and NSFC under 11474179, and the Tsinghua University Initiative Scientific Research Program.

{\bf Author Contributions} C.~L., S.~S., and S.~T.\  performed the experiments. C.~L., S.~T., and J.~T.\ analysed the data. All authors contributed to the understanding the spectra.  Z.\ Y., S.\ T., and S.\ Z.\ developed the two-channel model of the dynamics.  All authors contributed to the preparation of the manuscript.

{\bf Additional Information} Supplementary Information is available online, discussing the two-channel model and fitting systematics. Correspondence and requests for materials should be addressed to J.~T.\ and S.~Z.

{\bf Competing Financial Interests} The authors declare no competing financial interests.

\clearpage

\setcounter{figure}{0}
\setcounter{equation}{0}
\renewcommand\thefigure{S\arabic{figure}}
\renewcommand\theequation{S\arabic{equation}}

\section{Supplementary Material}

{\bf Connection between closed-channel molecules and p-wave contacts}

Let the creation operators for molecules and fermions at position ${\bf R}$ be $\phi^\dagger({\bf R})$ and $\psi^\dagger(\vec R)$, respectively, and the internal wave function of the molecules be given by $g({\bf r})\equiv g(r)Y_{1m}(\hat{r})$, with ${\bf r}$ the relative coordinate. Then the interaction Hamiltonian that converts scattering fermions to molecules can be written as
\begin{align}
H_{\rm int}\sim g\int d{\bf r}\,d{\bf R}\,\phi^\dagger({\bf R})g({\bf r})\psi({\bf R}+\frac{{\bf r}}{2})\psi({\bf R}-\frac{{\bf r}}{2})+{\rm H.c.}
\end{align}
where $g$ is the (un-renormalised) coupling constant. Transforming to Fourier space and noticing that $g({\bf r})$ is an odd function, which requires its Fourier transform to be linearly proportional to $kY_{1m}(\hat{k})$, the effective coupling Hamiltonian can be written as \cite{Ohashi2005,Gurarie:2005it,Cheng:2005kv}
\begin{align}
H_{\rm int}=\frac{\bar{g}}{\sqrt{V}}\sum_{{\bf k},{\bf p}}kY_{1m}(\hat{k})b^\dagger_{\bf p}a^\pdagger_{{\bf p}/2+{\bf k}}a^\pdagger_{{\bf p}/2-{\bf k}}+{\rm H.c.}\label{effh}
\end{align}
where $b^{\dagger}_{\bf p}$ ($a^{\dagger}_{\bf p}$) creates a molecule (fermion) with momentum ${\bf p}$, and $\bar{g}$ is the coupling constant. We shall relate the latter to the scattering volume $v_m$ and also the effective range $R_m$, for the scattering channel characterised by the magnetic quantum number $m$.  The free Hamiltonian, on the other hand, can be written as 
\begin{align} \label{h0}
H_0=\sum_{\bf k}\epsilon({\bf k})a_{\bf k}^\dagger a^\pdagger_{\bf k}+\sum_{\bf p} E({\bf p})b_{\bf p}^\dagger b^\pdagger_{\bf p},
\end{align}
where $E({\bf p})={\bf p}^2/(4M)+\nu_m(B)$, where $\nu_m(B)=\delta\mu_m(B-B_{0,m})+\mbox{const.}$ is the detuning of the closed-channel molecule. $\delta\mu_m$ is the magnetic moment difference between the open-channel scattering state and the closed-channel $p$-wave dimer.  

The connection of $\bar{g}$ and $\nu_m$ to $v_m$ and $R_m$ can be established by calculating the fermionic scattering $T$-matrix. The two-body Schr\"{o}dinger equation can be written as
\begin{align}
(E-H_0)\Psi=H_{\rm int}\Psi,
\end{align}
and can be solved conveniently as $\Psi=\Psi_0+(E-H_0)^{-1}H_{\rm int}\Psi\equiv \Psi_0+(E-H_0)^{-1}T\Psi_0$, where $\Psi_0$ is the asymptotic state in the absence of interaction and we have defined the scattering $T$-matrix, $T_m$. Developing, as usual, the perturbative series for $T_m$, one obtains, in the centre-of-mass frame
\begin{align}
\frac{1}{T_m(E)}=E-\nu_m(B)-\frac{2\bar{g}^2}{V}\sum_{{\bf k}}\frac{k^2|Y_{1m}(\hat{k})|^2}{E+i0^+-k^2/M},
\end{align}
where we note that the factor of $2$ arises from the exchange symmetry of the identical fermions. To evaluate the integral, we need to impose an ultra-violet cutoff $\Lambda$ in momentum. By further matching the relation with $1/T_m(E=\hbar^2 q^2/M)\sim-1/(v_m q^3)-1/(R_m q)-i$, we find
\begin{align}
\frac{1}{v_m} &=-\nu_m\frac{8\pi^2\hbar^2}{M\bar{g}^2}+\frac{2\Lambda^3}{3\pi},\label{rev}\\
\frac{1}{R_m} &=\frac{8\pi^2\hbar^4}{M^2\bar{g}^2}+\frac{2\Lambda}{\pi}.\label{reR}
\end{align}
These two equations establish the relation between the two parameters in the Hamiltonian $\{\bar{g},\nu_m\}$ and  the physical parameters $\{v_m,R_m\}$. 

To obtain the number of closed channel molecules, it is simplest to use the Hellmann-Feynman theorem, by taking the derivative with respect to $\nu_m$. That is
\begin{align}
N_{\dd,m}=\frac{\partial E}{\partial \nu_m}=\frac{\partial E}{\partial v_m^{-1}}\frac{\partial v_m^{-1}}{\partial \nu_m}.
\end{align}
Using the thermodynamic relation for $C_v$, derived from equation~(\ref{eq:thermoID}) in the main text, and the fact that close to resonance, $v^{-1}_m=-(B-B_{0,m})/(\Delta_m v_{m}^{\rm bg})$, we find that the number of dimers $N_{\dd,m}$ can be written as \cite{Yoshida2015}
\begin{align}
N_{\dd,m}=\frac{\hbar^2}{2 M\delta\mu^{(m)} \Delta_m v_{m}^{\rm bg}}C_{v,m}.
\end{align}
In fact, close to resonance, since $R_m\sim1/\Lambda\sim r_0$ with $r_0$ being the size of the closed channel molecules, the coefficient is just of order $1/R_m$. Namely,
\begin{equation}
C_{v,m}\sim 2 R_m N_{\dd,m} \quad  \mbox{for} \quad v_m \to \pm \infty
\end{equation}
such that $C_{v,m}$ is directly proportional to the number of $p$-wave dimers in the closed channel, and the proportionality constant is of the order of the effective range $R_m>0$. As a result, monitoring the value of $C_{v,m}$ through rf spectroscopy is equivalent to monitoring the number of dimers in the closed channel. 

Likewise, combining the thermodynamic relations with equations~(\ref{rev}) and (\ref{reR}), we find
\begin{align}\nonumber
-\frac{\hbar^2 C_{R,m}}{2 M}=&\left.\frac{\partial E}{\partial R_m^{-1}}\right|_{v_m}\\\nonumber
=&\left.\frac{\partial E}{\partial \nu_m}\right|_{\bar{g}}\left.\frac{\partial \nu_m}{\partial  R_m^{-1}}\right|_{v_m}+\left.\frac{\partial E}{\partial \bar g^{-2}}\right|_{\nu_m}\left.\frac{d\bar g^{-2}}{\partial R_m^{-1}}\right|_{v_m}\nonumber\\
=&-\frac{M^2\bar g^2}{8\pi^2\hbar^4}\left(\nu_m N_{\dd,m}+\frac12\langle H_\text{int}\rangle\right),
\end{align}
where $H_\text{int}$ is given in equation~(\ref{effh}). While the first term is related to the number of closed channel molecules, the later term involves the expectation value
\begin{align}
\left\langle b^\dagger_{\bf p}a^\pdagger_{{\bf p}/2+{\bf k}}a^\pdagger_{{\bf p}/2-{\bf k}}\right\rangle
\end{align}
which describes the atom pair-molecule coherence.

The model combining equations~(\ref{effh}) and (\ref{h0}) has been used to study $p$-wave superfluidity in Fermi gases \cite{Ohashi2005,Gurarie:2005it,Cheng:2005kv}. At zero temperature, at the mean field level, if condensation of molecules occurs only in the $m$th partial wave, one can show $C_{R,m}= 2 \mu C_{v,m} M/\hbar^2$, where $\mu$ is the chemical potential conjugate to the total number of fermions. The difference between $2\mu C_{v,m} M/\hbar^2$ and $-2 R_m^2 N_{\mathrm{d},m}/v_m$ is the contribution to $C_{R,m}$ from $\langle H_\text{int} \rangle$ in this case.

\smallskip
{\bf Decay of quasi-bound p-wave dimers}

\begin{figure}[tb!]
\includegraphics[width=0.45\textwidth]{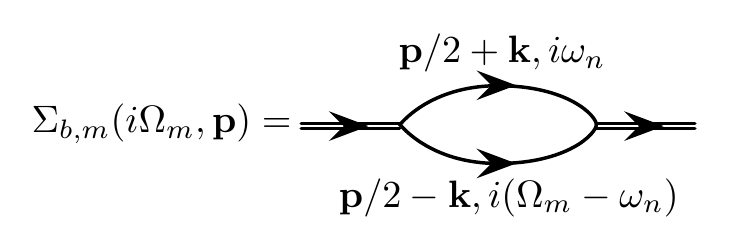}
\caption{{\bf Lowest order self-energy diagram for the dimer-dimer propagator.} The vertex is given by $\bar{g}kY_{1m}(\hat{k})/\sqrt{V}$. The double line is for dimer-dimer propagator and single lines are for fermion propagators.}
\label{figS3}
\end{figure}

Our next task is to investigate how a $p$-wave dimer above resonance ($E_{\dd,m}>0$) decays into scattering fermions. It is simplest to compute the self-energy of the dimer propagator $\Sigma_{b,m}(i\Omega_n,{\bf p})$. The lowest order diagram is given in Fig.~\ref{figS3}.
 
Explicitly, when analytically continued to real frequency $\Omega$,
\begin{align}
\Sigma_{b,m}(\Omega+i0^+,{\bf p})=\frac{2\bar{g}^2}{V}\sum_{\bf k}k^2|Y_{1m}(\hat{k})|^2\frac{f(\xi_{{\bf p}/2+{\bf k}})+f(\xi_{{\bf p}/2-{\bf k}})-1}{\hbar\Omega+i0^+-\xi_{{\bf p}/2+{\bf k}}-\xi_{{\bf p}/2-{\bf k}}},
\end{align}
where $\xi_{\bf k}=\hbar^2k^2/2M-\mu$, and $f(x)$ is the Fermi distribution function, describing the effects of the Pauli exclusion principle. In the two-body case (vacuum scattering), $\mu=0$ and the imaginary part of the self-energy is given by 
\begin{align}
\hbar\gamma_{m}(\Omega,{\bf p}) &\equiv 2{\rm Im}\Sigma_{b,m}(\Omega+i0^+,{\bf p})\nonumber \\
&=\frac{M\bar{g}^2}{4\pi^2\hbar^2}\left(\frac{M\Omega}{\hbar}-\frac{p^2}{4}\right)^{3/2}.
\end{align}
For ${\bf p}={\bf 0}$ and when $\hbar\Omega=E_{\dd,m}>0$, the imaginary part is non-zero and is given by
\begin{align}\label{eq:gammam}
\hbar\gamma_m\equiv \hbar\gamma_m(E_{\dd,m},{\bf 0})&=\frac{M\bar{g}^2}{4\pi^2\hbar^2}\left(\frac{ME_{\dd,m}}{\hbar^2}\right)^{3/2}\nonumber\\
&=2\,\frac{\hbar^2 R_m}{M}\left(\frac{ME_{\dd,m}}{\hbar^2}\right)^{3/2}.
\end{align}
In the main text this relation is given as $\gamma_m = 2 \sqrt{M}R_m E_{\dd,m}^{3/2}/\hbar^2$.

\smallskip
{\bf Analysis of the spectroscopic data}

\begin{figure}[tb!]
\includegraphics[width=0.5\textwidth]{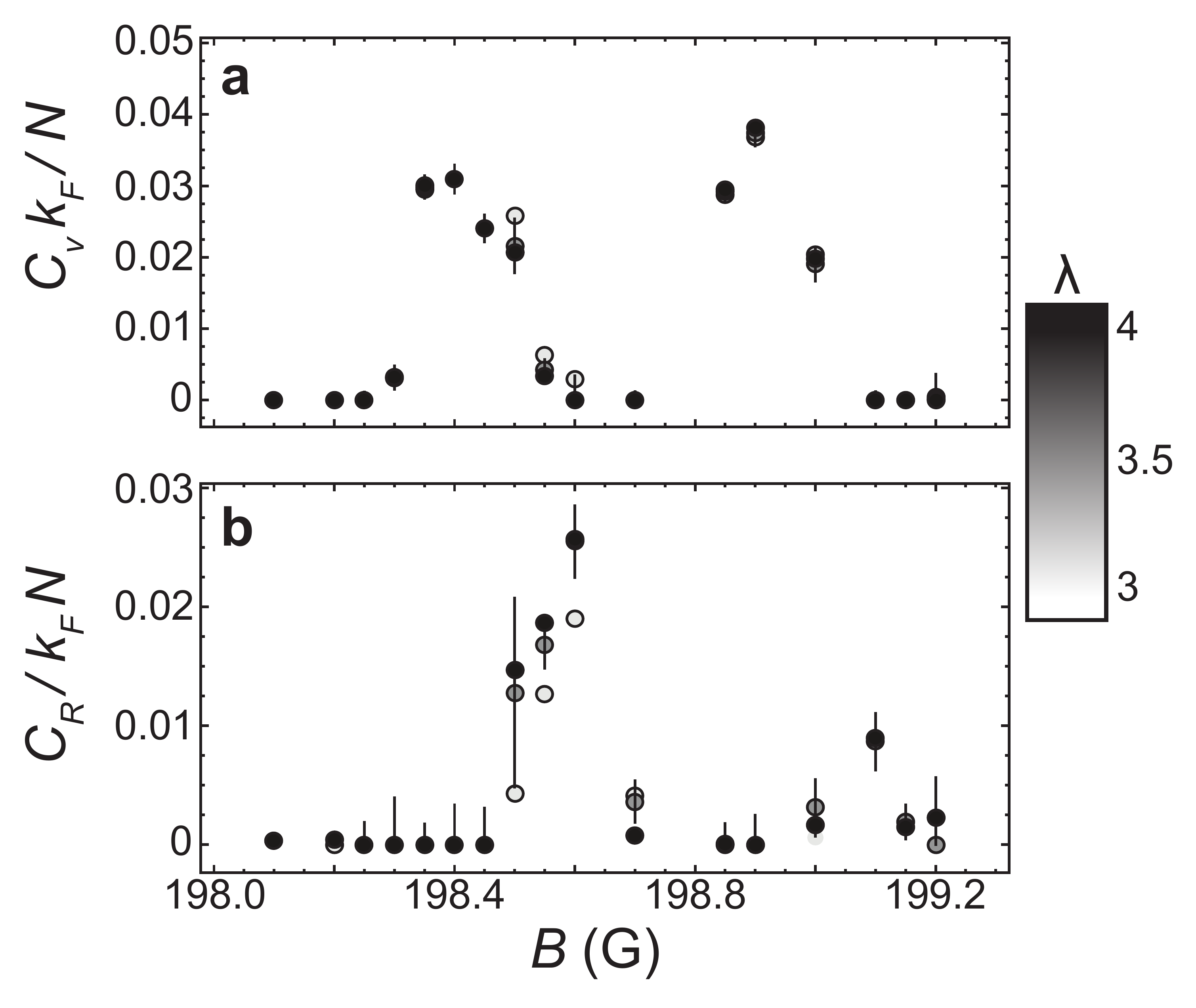}
\caption{{\bf Effect of cutoff in analysis of rf spectra.} The $p$-wave contacts {\bf a,} $C_v$ and {\bf b,} $C_R$ for various values of $\lambda$ plotted as a function of magnetic field. The data plotted here is analysed with $\lambda=3.0\,,3.5\,,4.0$. The error bars are determined in the fit for $\lambda=3.5$.}
\label{fig:lambdadep}
\end{figure}

{\textit{Selection of the high-frequency tail of the rf spectrum.}}
Since the contact only describes the functional form of rf spectra in the frequency range $\widetilde{\omega} \gg 1$, we analyse only data above some cut-off frequency $\widetilde{\omega} \ge \lambda$, typically $\lambda = 3.5$. 

There are two magnetic field regimes in which the fit results are insensitive to $\lambda$. For $\delta{B}_m\lesssim 0.1\,$G, the spectra are dominated by $C_v$; whereas for $\delta{B}_m \gtrsim 0.2\,$G, the spectra are dominated by $C_R$. However for values of $\delta{B}_m$ in between these two limits, the relative weight of the $\widetilde\omega^{-1/2}$ and $\widetilde\omega^{-3/2}$ components of the spectra is sensitive to $\lambda$. 

We report data with $\lambda = 3.5$ in Fig.~\ref{fig:contacts} of the main text where the error bars show statistical errors from the fit. We repeat our analysis for $\lambda=3,\,3.5,\,4$ as shown in Fig.~\ref{fig:lambdadep}. There is some scatter in the fitted values of $C_v$ and $C_R$, although it is roughly the same magnitude as the error bars from the fit. The reduced scatter in the data at the $m=z$ resonance is due to increased sampling of the spectra. The general behaviour of $C_v$ and $C_R$ are, however, not affected by the values of the chosen cutoff. With improved signal at large detuning we could minimise this region that is sensitive to the choice of $\lambda$.

\begin{figure}[tb!]
\includegraphics[width=0.6\textwidth]{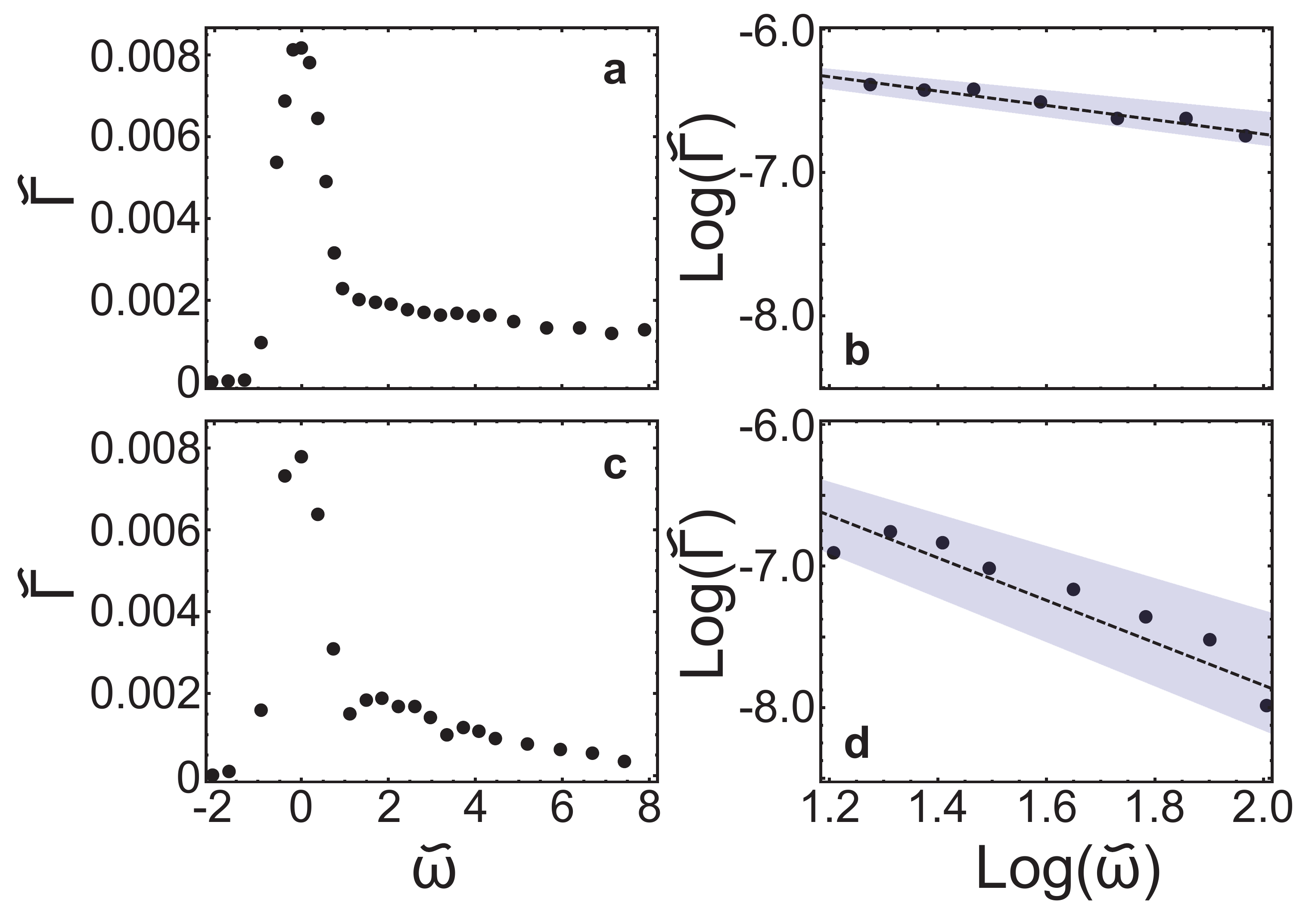}
\caption{{\bf Determination of the power law in two limits of} $\mathbf{\delta B}${\bf .} {\bf a,} At $\delta{B}_z=0.05$ G the spectrum is dominated by $C_v$. {\bf b,} A log-log plot of the high frequency tail. The shaded region shows the fitted power law $-0.43\pm0.06$ and the dashed line shows the prediction of $-0.5$. {\bf c,} At $\delta{B}_{xy}=0.3$ G the spectrum is dominated by $C_R$. {\bf d,} When plotted on a log-log plot the fitted power law with slope $-1.35\pm0.21$ (shaded region) is consistent with the prediction of $-1.5$ (dashed line).}
\label{fig:powerlaw}
\end{figure}

{\textit{Determination of the asymptotic power law.}}
At each field value we fit equation~(\ref{eq:Irf}) from the main text to the high-frequency tail of our spectroscopic measurements of $\widetilde{\Gamma}$. The high-frequency cut-off is chosen as discussed above and the prefactor to each power law is left as a free parameter. If we choose only one power law, setting the other prefactor to zero, the fits in the regime $0.1\lesssim\delta{B}\lesssim0.2$ are poorly constrained. Further evidence that both power laws are required to fully describe our data is motivated by Fig.~\ref{fig:contacts} in the main text. There are two regimes in $\delta{B}$ where either only $C_v$ or $C_R$ contribute to the spectrum. For $\delta{B}<0.1$ G $C_v$ is dominant and therefore only the $\widetilde{\omega}^{-1/2}$ power law scaling is present in the spectrum. This is shown in Fig.~\ref{fig:powerlaw}a at $\delta{B}_z=0.05$ G. Fitting the high-frequency tail of this spectrum with a function, $f(\widetilde{\omega})=A\widetilde{\omega}^{\eta}$, we find an exponent $\eta=-0.43\pm0.06$. This power law is made more apparent when plotting the high frequency tail of the spectrum in a log-log plot. The power law appears as a constant slope shown as the shaded region in Fig.~\ref{fig:powerlaw}b and compared to the theoretical slope of $-0.5$ shown as the dashed line.

In contrast, for $\delta{B}>0.2$ G $C_R$ is the dominant contribution to the spectrum. Experimentally, this appears as a change in the fitted value of $\eta$. At $\delta{B}_{xy}=0.3$ G we find $\eta=-1.35\pm0.21$ from the data shown in Fig.~\ref{fig:powerlaw}c. Qualitatively, the difference in power law can be seen when comparing Fig.~\ref{fig:powerlaw}a and c. The weight at high frequency in Fig.~\ref{fig:powerlaw}c goes to zero much quicker as $\widetilde{\omega}$ is increased than in Fig.~\ref{fig:powerlaw}a. When plotted on a log-log plot (see Fig.~\ref{fig:powerlaw}d) the high frequency behaviour is clearly different. The fitted power law (shaded region) appears as a slope consistent with the theoretical value of $-1.5$ (dashed line). $s$-wave interactions would show a similar spectrum, but could only appear due to an unintentional admixture of incoherent atoms in state $\ket{1}$. We estimate this background to be less than 2\% of the observed signal.

\smallskip
{\bf Analysis of the momentum distribution data}

\begin{figure}[b!]
\includegraphics[width=0.3\textwidth]{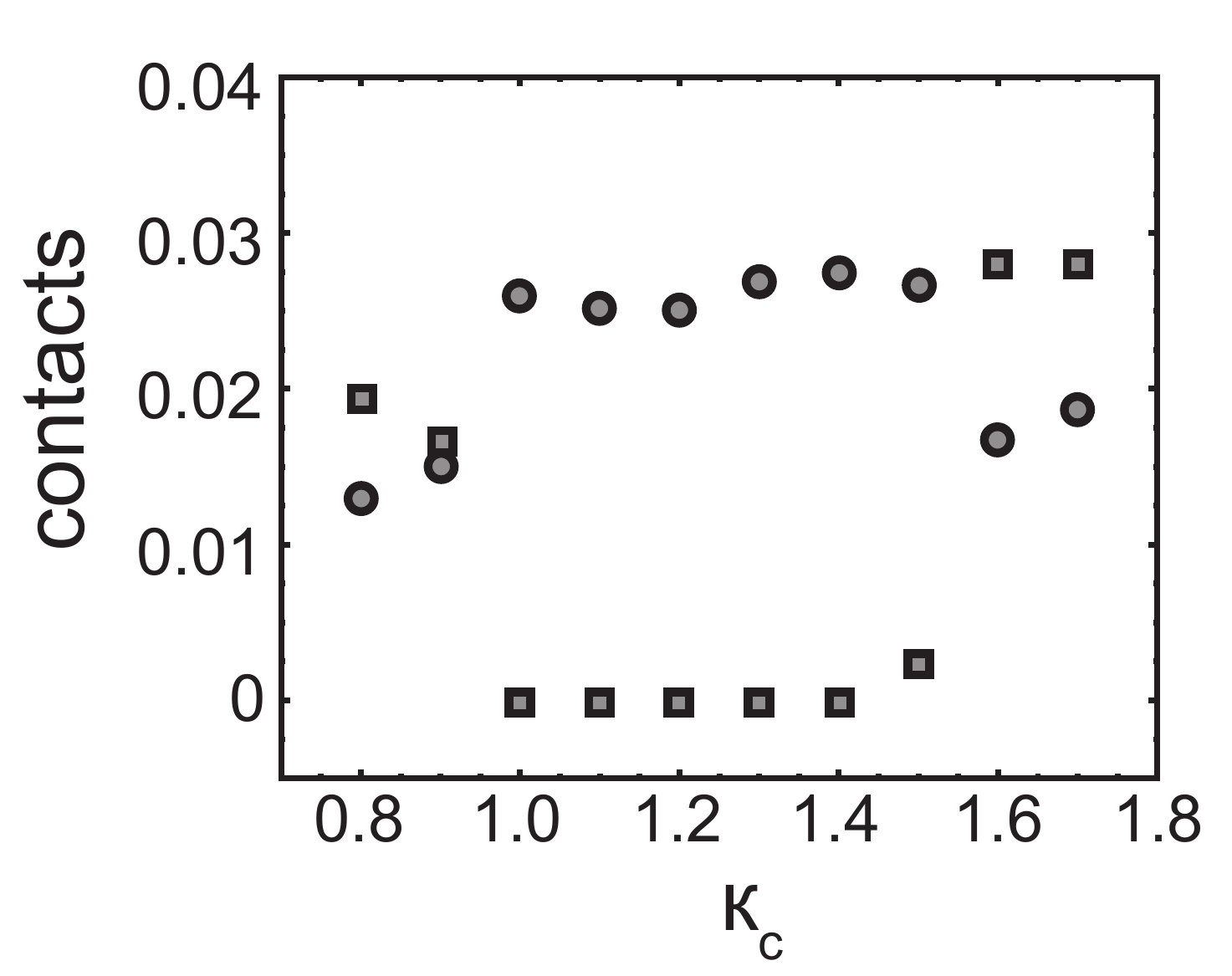}
\caption{{\bf Effect of low-momentum cutoff in analysis.} The $p$-wave contacts $C_vk_{\rm{F}}/N$ (circles) and $C_R/k_{\rm{F}}N$ (squares) extracted from the momentum distribution at $\delta B_{xy}=+0.15$ G as a function of momentum cut-off. A plateau in the fitted values appears around $\kappa_c\approx1.2$. }
\label{fig:momcutoff}
\end{figure}

To improve the signal-to-noise in this measurement we perform several forms of averaging. Firstly, we average approximately 40 line-of-sight images. To perform this average we determine the centre of each atom cloud through a two-dimensional gaussian fit. We then overlay these centres to construct an average image. Secondly, we perform a radial average. This serves the dual purpose of improving signal and averaging out noise. Using the centre determined in the previous averaging, we calculate the radial distance to each pixel, $\kappa=\sqrt{k_x^2+k_y^2}/k_{\rm{F}}$. We then calculate the average of all pixels at the same radius. Finally, we bin pixels to construct the distribution shown in Fig.~\ref{fig:spectrum}b, which has sufficient signal to perform a two parameter fit at large momentum.

At each magnetic field the momentum distribution measured is highly dependent on the atom number which varies with the detuning from resonance (as the loss rate changes). As such, we independently choose a momentum cut-off $\kappa_c$ for each data set.
To determine the momentum cut-off we perform a power law fit using equation~(\ref{eq:nk}) from the main text and vary the value of $\kappa_c$. The result is shown in Fig.~\ref{fig:momcutoff} for data at $\delta B_{xy}=0.15$ G. As can be seen, a plateau occurs over a range of cut-off values near $\kappa_c\approx1.2$. We typically choose the mean value of this plateau as the cut-off for a given data set.

\begin{figure}[b]
\includegraphics[width=0.45\textwidth]{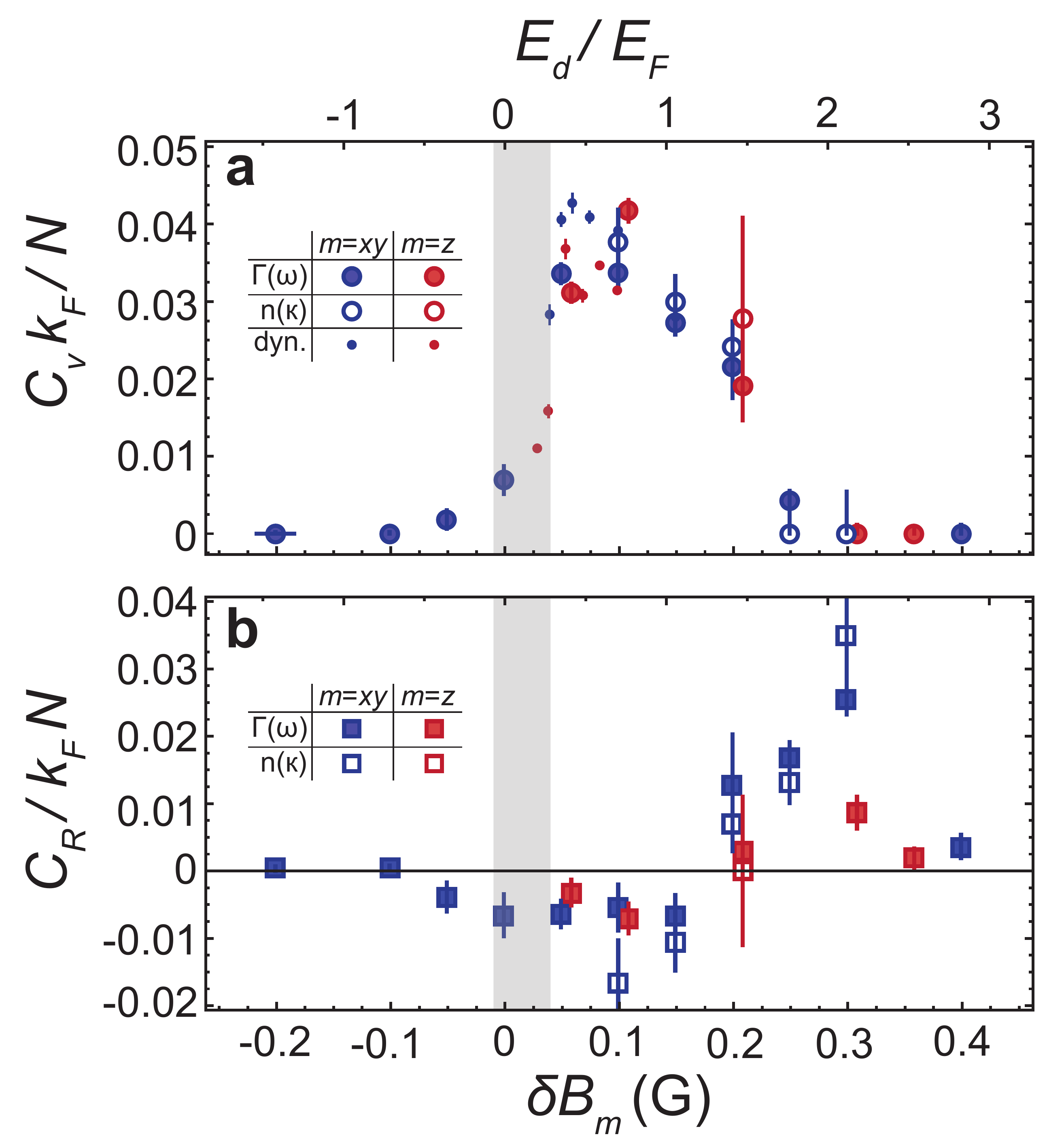}
\caption{{\bf The $p$-wave contacts allowing negative $\bm{C_R}$.} {\bf a,} $C_v$ and {\bf b,} $C_R$ for various values of $\delta{B}$ with $C_v>0$ and $C_R$ unconstrained. Negative values of $C_R$ are observed near resonance.}
\label{fig:negCR}
\end{figure}

\begin{figure}[t]
\includegraphics[width=0.4\textwidth]{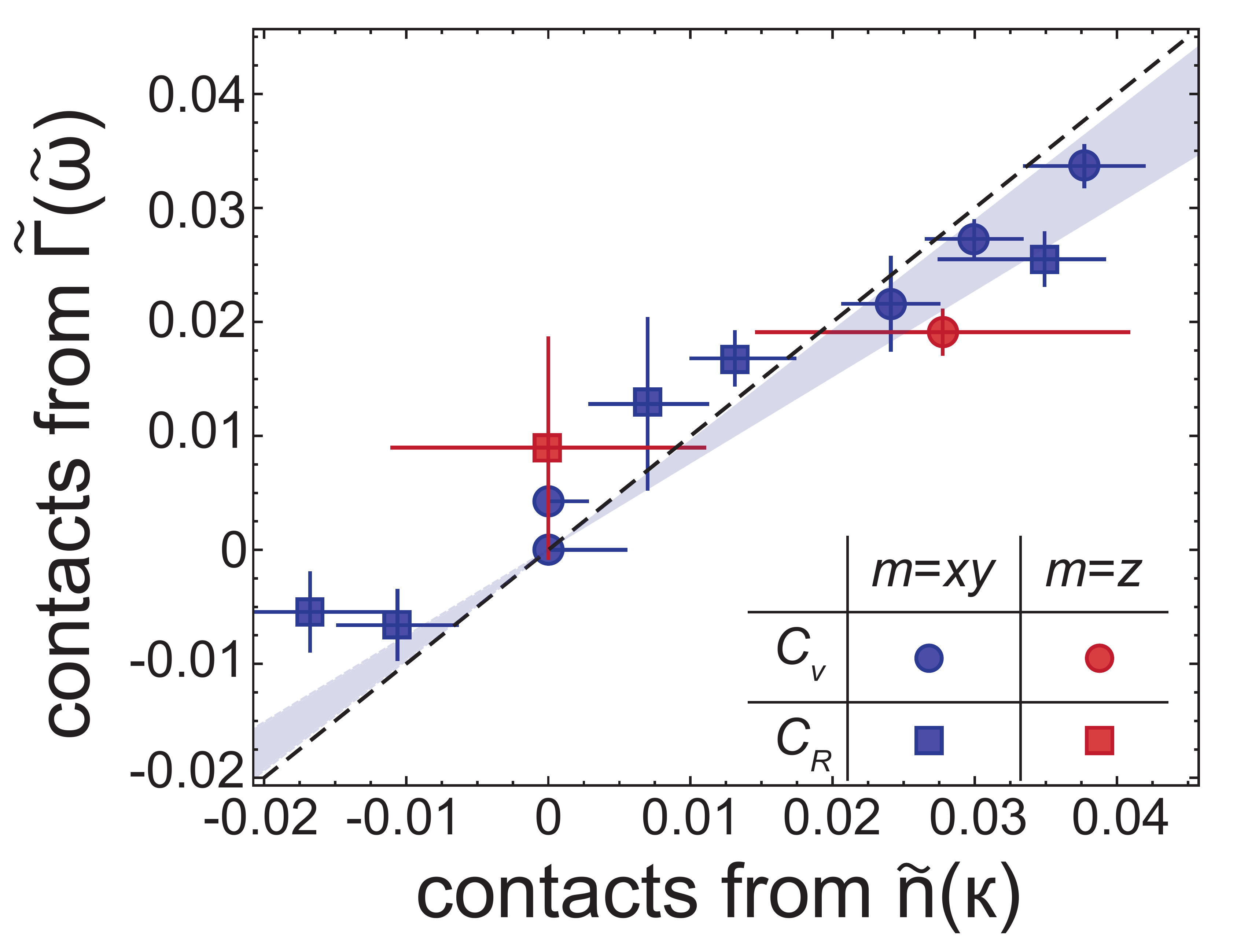}
\caption{{\bf Comparison of the contacts, allowing negative $\bm{C_R}$.} A comparison of $C_v k_\mathrm{F}/N$ or $C_R / k_\mathrm{F} N$ from both the momentum distribution and spectroscopy. Here, $C_R < 0$ is allowed, unlike in the main text.}
\label{fig:comp}
\end{figure}

\smallskip
{\bf Systematic effect of allowing negative C$_R$}

The values returned in the fits of our data with equations~(\ref{eq:nk}) and (\ref{eq:Irf}) from the main text are sensitive to noise, atom loss at high energy, and small offsets due to imperfect background subtraction. These systematics combine to provide an uncertainty in the values of $C_R$, especially in the regions dominated by a large $C_v$. Figure~\ref{fig:negCR} shows values extracted for the contacts with $C_R$ unconstrained. We find small negative values for $C_R$ near resonance that are not consistent with zero.

Our constrained fits in the main text are motivated in part by the fact that a small offset added to the spectrum can result in values of $C_R$ consistent with zero. This offset is typically $\sim50$ atoms, consistent with the scatter in our number counting when no atoms are present. Further motivation for performing constrained fits comes from repeating this analysis on high-frequency tails simulated with Gaussian noise matching experimental conditions. For the maximum values of $C_v$ observed in this experiment we cannot statistically distinguish between a negative value of $C_R$ and a small positive offset.

We compare the results of these unconstrained fits for data from both the momentum distribution and spectroscopy in Fig.~\ref{fig:comp}. A linear fit constrained to pass through the origin finds a slope of $0.86(9)$ (grey area), where uncertainty is statistical; compared to $0.96(7)$ in the main text.

\smallskip
{\bf Rate equations near a p-wave resonance}

In the two-channel model described above, the dynamics of $C_v$ can be understood by considering how the dimer population changes as a function of time. Let the momentum distribution of the dimers be given by $n_m({\bf p})$ and assume that the conversion rate between dimers and scattering fermions occurs at the same rate as that given in vacuum, $\gamma_m$, irrespective of the asymptotic energy of the scattering fermions. This is a good approximation when the p-wave dimers are only slightly above the threshold $E_{\dd,m}\gtrsim 0$, or very close to the $p$-wave resonance. In this case, due to energy and momentum conservation, if the centre-of-mass momentum of the dimer is ${\bf p}$, then the incoming scattering fermion must have momentum ${\bf p}/2+{\bf k}$ and ${\bf p}/2-{\bf k}$, with $|{\bf k}|=k_0$ and $\hbar^2k_0^2/M=-E_{\dd,m}$. The magnitude of ${\bf k}$ is fixed by the dimer binding energy, while its direction can be arbitrary. In the vacuum case, the angular average gives rise to the decay rate of the $p$-wave dimer, $\gamma_m$. Thus we can write down the {\em phenomenological} rate equations for the populations of dimers and fermions, 
\begin{align}
\frac{d n_m({\bf p})}{dt}=\gamma_m\sum_{|{\bf k}|=k_0}\Big[\overline{f_{{\bf p}/2+{\bf k}}f_{{\bf p}/2-{\bf k}}}(1+n_m({\bf p}))-\overline{(1-f_{{\bf p}/2+{\bf k}})(1-f_{{\bf p}/2-{\bf k}})}n_m({\bf p})\Big],
\end{align}
where $\overline{ff}$ means the angular average over ${\bf k}$. The first term describes the conversion of two fermions into a dimer, while the second term describes the opposite process in which a dimer disassociates into two fermions, corrected with Fermi and Bose statistics. 

At the beginning of the dynamics we assume that there is a Fermi sea of Fermi momentum $k_{\rm F}$. When $E_{\dd,m}>2E_{\rm F}$ ($k_0>k_{\rm F}$) with $E_{\rm F}=k_{\rm F}^2/2M$, the conversion into the the dimers from the fermions cannot happen, since energy and momentum conservation cannot be simultaneously maintained. On the other hand, in the limit $k_0/k_{\rm F}\ll 1$, i.e., that the dimer is only slightly above the threshold, we can write
\begin{align}\nonumber
\frac{d n_m({\bf p})}{dt}&=\gamma_m\Big[f^2_{{\bf p}/2}(1+n_m({\bf p}))-(1-f_{{\bf p}/2})^2n_m({\bf p})\Big]\\
&=\gamma_m[f^2_{{\bf p}/2}-(1-2f_{{\bf p}/2})n_m({\bf p})]\label{dnmt}.
\end{align}
In the experiments, we are interested in the total number of dimers, which is related to $C_{v,m}$. Summing over ${\bf p}$ on both side of equation~(\ref{dnmt}), one finds
\begin{align}
\frac{d N_{\dd,m}}{dt}=&\gamma_m\sum_{\bf p}[f^2_{{\bf p}/2}-(1-2f_{{\bf p}/2})n_m({\bf p})]\\\nonumber
=&-\gamma_m N_{\dd,m}+\gamma_m\sum_{\bf p}f^2_{{\bf p}/2}+2\gamma_m\sum_{\bf p}f_{{\bf p}/2}n_m({\bf p}).
\end{align}
The first term describes the vacuum decay of a dimer, proportional to $N_{\dd,m}$. The second term describes the conversion of two fermions into a dimer, which when summed over ${\bf p}$, gives $8\gamma_m N_\ff$ at zero temperature. The last term arises due to the Pauli principle, which inhibits the decay of dimers into fermions when the final states are already occupied and therefore reduces the apparent dimer decay rate. This term is more complicated to handle, and we combine it with the second term to be described by an empirical association rate $A_m$. 

We can repeat the same steps for the density of scattering fermions and arrive at a set of coupled differential equations for $N_{\dd,m}$ and $N_\ff$:
\begin{eqnarray}\label{eq:rateeqNd}
  \frac{d N_{\dd,m}}{dt}&=&-\gamma_m N_{\dd,m}+A_m N_\ff  \\
  \label{eq:rateeqNf}
  \frac{d N^{}_\ff}{dt}&=&\sum_m \left(2\gamma_m N_{\dd,m}-2A_m N_\ff\right)\,\label{eq:rateeqNN}
\end{eqnarray}
which conserve the total number of fermions $N = N_\ff + 2\sum_{m} N_{\dd,m}$, and where the sum includes those closed channels relevant for the respective resonance. The above arguments indicate in our low temperature experiment $A_m=0$ when $E_{\dd,m}>2E_{\rm F}$, and $A_m\sim 8\gamma_m$ when $E_{\dd,m}$ is only slightly above the threshold.

It is worth emphasizing that the scaling of the association term with $N_\ff$ is a result of the degenerate open-channel Fermi sea. In the Boltzmann regime, a similar calculation predicts the intuitive $N_\ff^2$ scaling for a two-body process: the association part $A_mN_\ff$ in equations~(\ref{eq:rateeqNd}) and (\ref{eq:rateeqNN}) is replaced by 
\begin{equation}
A_m N_\ff \rightarrow \gamma_m \frac{N_\ff^2}{V} \left(\frac{4\pi\hbar^2}{Mk_BT}\right)^{3/2} \exp{\left[ \frac{\hbar^2R_m}{v_mMk_BT} \right]}.
\end{equation}
The physical reason for this scaling is that the dimer association is the reverse process of dimer decay, which in vacuum would occur at $\gamma_m$. The ratios are determined by the phase space density of the fermions that pair into dimers. In a thermal cloud, the ratio is proportional to $n \Lambda^3$, where $\Lambda$ is the de Broglie wavelength and $n$ is the local density. For a degenerate Fermi cloud, the initial rate saturates to $8 \gamma$ at $1/v=0$ and $T=0$. Thus the observed magnitude of $A_m$ reflects the degenerate nature of the cloud throughout the fast dynamics.

{\textit{Empirical atom-loss term.}}
We consider two different inelastic collisions that lead to a loss of atoms from the trap: collisions between two dimers $(L_{\dd\dd})$ and between a fermion and a dimer $(L_{\ff\dd})$. Including these processes with respective loss rates $L$, equations~(\ref{eq:rateeqNd},\ref{eq:rateeqNf}) take the form
\begin{eqnarray}\label{eq:rateeqNdloss}
   \dot{N}_{\dd,m} &=& -\gamma_m N_{\dd,m} + A_m N_\ff - L^{(m)}_{\ff\dd}\,N_\ff  N_{\dd,m} - \sum_{m'} L_{\dd\dd}^{(m,m')}  N_{\dd,m} N_{\dd,m'}\\
   \dot{N}_\ff &=& -2\sum_m \Big(A_m N_\ff - \gamma_m  N_{\dd,m}  + \tfrac{1}{2} L_{\ff\dd}^{(m)}\,N_\ff  N_{\dd,m}\Big)\,. \nonumber \\
&& \label{eq:rateeqNfloss}
\end{eqnarray}
Near the $p_{z}$ resonance, equations~(\ref{eq:rateeqNdloss},\ref{eq:rateeqNfloss}) provide two differential equations that we use to model $C_v(t) \propto N_{\dd,z}(t)$ and $N(t)$. Near the $p_{xy}$ resonance ($m=x, y$), we have three coupled differential equations for $N_\ff$, $N_{\dd,x}$ and $N_{\dd,y}$. These can be reduced to a pair of equations for $N_\ff$ and $N_{\dd,xy} \equiv N_{\dd,x} + N_{\dd,y}$ if one assumes $A_x = A_y$, $\gamma_x = \gamma_y$, $L_{\ff\dd}^{(x)} = L_{\ff\dd}^{(y)}$, and $L_{\dd\dd}^{(x)} = L_{\dd\dd}^{(y)}$. Our three-dimensional geometry ensures cloud radii that exceed $(v_m)^{-3}$ and $R_m$, so that all these conditions are well fulfilled. We can therefore use the same set of two equations for the $p_z$ and the $p_{xy}$ resonances, where we interpret $A_{xy} = A_x + A_y$ in the latter case.

\begin{figure}[tb!]
\includegraphics[width=0.45\textwidth]{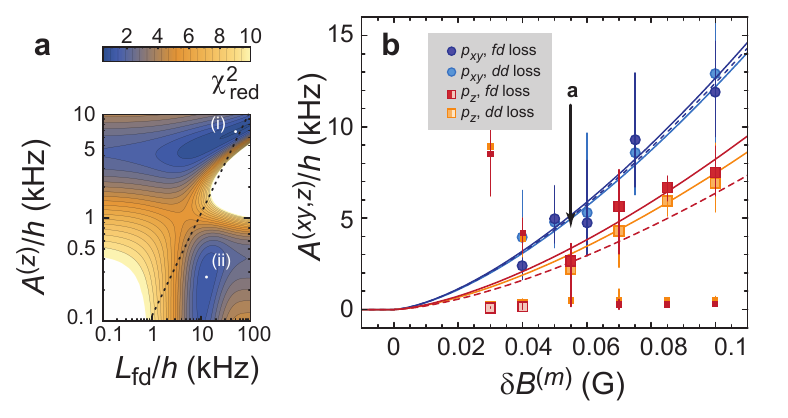}
\caption{{\bf Analysis of dynamics data.} {\bf a,} Typical $\chi_{\rm red}^2$ map (here: $B = 198.87\,{\rm G}$, corresponding to $\delta B_z = 0.08\,{\rm G}$) showing two local minima corresponding to the association-dominated ({\em i}) and the loss-dominated solutions ({\em ii}). The dashed line shows the condition $\tau_zL_{\ff\dd} = 1$. {\bf b,} Best fit values for the association rate at both resonances, and for both loss models. We also include the discarded local solutions for the $p_z$ resonance, which clearly violate the predicted proportionality to $\gamma_m \propto (\delta B_m^{3/2}$ (small symbols). Solid and dashed lines are as in Fig.~\ref{fig:dynamics}g in the main text.}
\label{figSDynamics}
\end{figure}

{\textit{Analysis of the dynamics data.}}
For each sampled field value, we fit the solution of equations~(\ref{eq:rateeqNdloss},\ref{eq:rateeqNfloss}) simultaneously to our dynamical measurements of $C_v(t) k_F/N(t)$ and $N(t)$. In this, we choose only one of the loss terms, setting the other loss rate to zero; varying more than one loss parameter leads to poorly constrained fits for our data. We furthermore assume a resonance position, which we use to calculate $\gamma_m$ through equation~(\ref{eq:gammam}). This leaves us with four fit parameters: the pre-factor $\alpha \equiv C_v/N_\dd$, the phenomenological association rate $A_m$, the respective loss rate $L$ and the initial atom number $N_0 \equiv N(t=0)$ which we need to include due to an unknown calibration factor in our state-selective imaging scheme. For a fixed grid of pairs $(A_m, L)$, we calculate the reduced chi-squared value $\chi_{\rm red}^2(A_m, L)$ for the optimal values of $\alpha$ and $N_0$. 

A typical distribution is shown in Fig.~\ref{figSDynamics}a for a field value near the $p_z$ resonance. We typically find two local minima with similar values of $\chi_{\rm red}^2$ for each set of dynamical data. One corresponds to the loss-dominated regime described in the main text, while the other would have the system reach a quasi-steady state. Figure~\ref{figSDynamics}b shows the association rates $A$ for either a pure fermion-dimer loss term (while $L_{\dd\dd} = 0$), or a pure dimer-dimer loss term (while $L_{\ff\dd} = 0$). Both loss models give compatible results. The result of the Fermi's golden rule calculation, $A_m \propto \gamma_m$, is employed to discard one solution for each time-resolved measurement (small symbols in Fig.~\ref{figSDynamics}b). The remaining best-fit values for $A_m$ outside the loss-dominated regime are then extrapolated to $A_m = 0$ to determine a new resonance position $B_0^{(m)}$. This new field value is assumed for a next iteration until the fit routine has converged.


\begin{thebibliography}{10}
\expandafter\ifx\csname url\endcsname\relax
  \def\url#1{\texttt{#1}}\fi
\expandafter\ifx\csname urlprefix\endcsname\relax\def\urlprefix{URL }\fi
\providecommand{\bibinfo}[2]{#2}
\providecommand{\eprint}[2][]{\url{#2}}

\bibitem{Tan2008:01}
\bibinfo{author}{Tan, S.}
\newblock \bibinfo{title}{Energetics of a strongly correlated {F}ermi gas}.
\newblock \emph{\bibinfo{journal}{Ann. Phys.}} \textbf{\bibinfo{volume}{323}},
  \bibinfo{pages}{2952 -- 2970} (\bibinfo{year}{2008}).

\bibitem{Tan2008:02}
\bibinfo{author}{Tan, S.}
\newblock \bibinfo{title}{Large momentum part of a strongly correlated {F}ermi
  gas}.
\newblock \emph{\bibinfo{journal}{Ann. Phys.}} \textbf{\bibinfo{volume}{323}},
  \bibinfo{pages}{2971 -- 2986} (\bibinfo{year}{2008}).

\bibitem{Tan2008:03}
\bibinfo{author}{Tan, S.}
\newblock \bibinfo{title}{Generalized virial theorem and pressure relation for
  a strongly correlated {F}ermi gas}.
\newblock \emph{\bibinfo{journal}{Ann. Phys.}} \textbf{\bibinfo{volume}{323}},
  \bibinfo{pages}{2987 -- 2990} (\bibinfo{year}{2008}).

\bibitem{Werner2009}
\bibinfo{author}{Werner, F.}, \bibinfo{author}{Tarruell, L.} \&
  \bibinfo{author}{Castin, Y.}
\newblock \bibinfo{title}{Number of closed-channel molecules in the {BEC-BCS}
  crossover}.
\newblock \emph{\bibinfo{journal}{Eur. Phys. J. B}}
  \textbf{\bibinfo{volume}{68}}, \bibinfo{pages}{401--415}
  (\bibinfo{year}{2009}).

\bibitem{Zhang2009}
\bibinfo{author}{Zhang, S.} \& \bibinfo{author}{Leggett, A.~J.}
\newblock \bibinfo{title}{Universal properties of the ultracold {F}ermi gas}.
\newblock \emph{\bibinfo{journal}{Phys. Rev. A}} \textbf{\bibinfo{volume}{79}},
  \bibinfo{pages}{023601} (\bibinfo{year}{2009}).

\bibitem{Braaten:2012gh}
\bibinfo{author}{Braaten, E.}
\newblock \bibinfo{title}{Universal relations for fermions with large
  scattering length}.
\newblock In \bibinfo{editor}{Zwerger, W.} (ed.) \emph{\bibinfo{booktitle}{The
  {BCS}-{BEC} Crossover and the Unitary {F}ermi Gas}},
  \bibinfo{pages}{193--231} (\bibinfo{publisher}{Springer},
  \bibinfo{address}{Berlin}, \bibinfo{year}{2012}),
\newblock \bibinfo{note}{{a}nd references therein.}

\bibitem{Werner2012}
\bibinfo{author}{Werner, F.} \& \bibinfo{author}{Castin, Y.}
\newblock \bibinfo{title}{{General relations for quantum gases in two and three
  dimensions: Two-component fermions}}.
\newblock \emph{\bibinfo{journal}{Phys. Rev. A}} \textbf{\bibinfo{volume}{86}},
  \bibinfo{pages}{013626} (\bibinfo{year}{2012}).

\bibitem{Werner:2012hy}
\bibinfo{author}{Werner, F.} \& \bibinfo{author}{Castin, Y.}
\newblock \bibinfo{title}{{General relations for quantum gases in two and three
  dimensions. II. Bosons and mixtures}}.
\newblock \emph{\bibinfo{journal}{Phys. Rev. A}} \textbf{\bibinfo{volume}{86}},
  \bibinfo{pages}{053633} (\bibinfo{year}{2012}).

\bibitem{Wild:2012fi}
\bibinfo{author}{Wild, R.~J.}, \bibinfo{author}{Makotyn, P.},
  \bibinfo{author}{Pino, J.~M.}, \bibinfo{author}{Cornell, E.~A.} \&
  \bibinfo{author}{Jin, D.~S.}
\newblock \bibinfo{title}{{Measurements of Tan{\textquoteright}s Contact in an
  Atomic Bose-Einstein Condensate}}.
\newblock \emph{\bibinfo{journal}{Phys. Rev. Lett.}}
  \textbf{\bibinfo{volume}{108}}, \bibinfo{pages}{145305}
  (\bibinfo{year}{2012}).

\bibitem{Olshanii:2003kn}
\bibinfo{author}{Olshanii, M.} \& \bibinfo{author}{Dunjko, V.}
\newblock \bibinfo{title}{{Short-Distance Correlation Properties of the
  Lieb-Liniger System and Momentum Distributions of Trapped One-Dimensional
  Atomic Gases}}.
\newblock \emph{\bibinfo{journal}{Phys. Rev. Lett.}}
  \textbf{\bibinfo{volume}{91}}, \bibinfo{pages}{090401}
  (\bibinfo{year}{2003}).

\bibitem{Combescot:2009gw}
\bibinfo{author}{Combescot, R.}, \bibinfo{author}{Alzetto, F.} \&
  \bibinfo{author}{Leyronas, X.}
\newblock \bibinfo{title}{{Particle distribution tail and related energy
  formula}}.
\newblock \emph{\bibinfo{journal}{Phys. Rev. A}} \textbf{\bibinfo{volume}{79}},
  \bibinfo{pages}{053640} (\bibinfo{year}{2009}).

\bibitem{Frohlich:2012ic}
\bibinfo{author}{Fr{\"o}hlich, B.} \emph{et~al.}
\newblock \bibinfo{title}{{Two-dimensional Fermi liquid with attractive
  interactions}}.
\newblock \emph{\bibinfo{journal}{Phys. Rev. Lett.}}
  \textbf{\bibinfo{volume}{109}}, \bibinfo{pages}{130403}
  (\bibinfo{year}{2012}).

\bibitem{Barth:2011ky}
\bibinfo{author}{Barth, M.} \& \bibinfo{author}{Zwerger, W.}
\newblock \bibinfo{title}{{Tan relations in one dimension}}.
\newblock \emph{\bibinfo{journal}{Ann. Phys.}} \textbf{\bibinfo{volume}{326}},
  \bibinfo{pages}{2544--2565} (\bibinfo{year}{2011}).

\bibitem{Weiss:2015gf}
\bibinfo{author}{Weiss, R.}, \bibinfo{author}{Bazak, B.} \&
  \bibinfo{author}{Barnea, N.}
\newblock \bibinfo{title}{{Nuclear Neutron-Proton Contact and the
  Photoabsorption Cross Section}}.
\newblock \emph{\bibinfo{journal}{Phys. Rev. Lett.}}
  \textbf{\bibinfo{volume}{114}}, \bibinfo{pages}{012501}
  (\bibinfo{year}{2015}).

\bibitem{DeMarco1999}
\bibinfo{author}{DeMarco, B.}, \bibinfo{author}{Bohn, J.~L.},
  \bibinfo{author}{Burke, J.~P.}, \bibinfo{author}{Holland, M.} \&
  \bibinfo{author}{Jin, D.~S.}
\newblock \bibinfo{title}{Measurement of $\mathit{p}$-wave threshold law using
  evaporatively cooled fermionic atoms}.
\newblock \emph{\bibinfo{journal}{Phys. Rev. Lett.}}
  \textbf{\bibinfo{volume}{82}}, \bibinfo{pages}{4208--4211}
  (\bibinfo{year}{1999}).

\bibitem{Regal2003}
\bibinfo{author}{Regal, C.~A.}, \bibinfo{author}{Ticknor, C.},
  \bibinfo{author}{Bohn, J.~L.} \& \bibinfo{author}{Jin, D.~S.}
\newblock \bibinfo{title}{Tuning $p$-wave interactions in an ultracold {F}ermi
  gas of atoms}.
\newblock \emph{\bibinfo{journal}{Phys. Rev. Lett.}}
  \textbf{\bibinfo{volume}{90}}, \bibinfo{pages}{053201}
  (\bibinfo{year}{2003}).

\bibitem{Zhang2004}
\bibinfo{author}{Zhang, J.} \emph{et~al.}
\newblock \bibinfo{title}{$p$-wave {F}eshbach resonances of ultracold
  $^{6}\mathrm{Li}$}.
\newblock \emph{\bibinfo{journal}{Phys. Rev. A}} \textbf{\bibinfo{volume}{70}},
  \bibinfo{pages}{030702} (\bibinfo{year}{2004}).

\bibitem{Schunck:2005cf}
\bibinfo{author}{Schunck, C.~H.} \emph{et~al.}
\newblock \bibinfo{title}{{{F}eshbach resonances in fermionic {Li}-6}}.
\newblock \emph{\bibinfo{journal}{Phys. Rev. A}} \textbf{\bibinfo{volume}{71}},
  \bibinfo{pages}{045601} (\bibinfo{year}{2005}).

\bibitem{Kallin:2012kx}
\bibinfo{author}{Kallin, C.}
\newblock \bibinfo{title}{{Chiral p-wave order in Sr$_2$RuO$_4$}}.
\newblock \emph{\bibinfo{journal}{Rep. Prog. Phys.}}
  \textbf{\bibinfo{volume}{75}}, \bibinfo{pages}{042501} (\bibinfo{year}{2012}).

\bibitem{Read:2000iq}
\bibinfo{author}{Read, N.} \& \bibinfo{author}{Green, D.}
\newblock \bibinfo{title}{{Paired states of fermions in two dimensions with
  breaking of parity and time-reversal symmetries and the fractional quantum
  Hall effect}}.
\newblock \emph{\bibinfo{journal}{Phys. Rev. B}} \textbf{\bibinfo{volume}{61}},
  \bibinfo{pages}{10267--10297} (\bibinfo{year}{2000}).

\bibitem{Levinsen:2007gy}
\bibinfo{author}{Levinsen, J.}, \bibinfo{author}{Cooper, N.~R.} \&
  \bibinfo{author}{Gurarie, V.}
\newblock \bibinfo{title}{Strongly resonant $p$-wave superfluids}.
\newblock \emph{\bibinfo{journal}{Phys. Rev. Lett.}}
  \textbf{\bibinfo{volume}{99}}, \bibinfo{pages}{210402}
  (\bibinfo{year}{2007}).

\bibitem{MajoranaReview}
\bibinfo{author}{Elliott, S.~R.} \& \bibinfo{author}{Franz, M.}
\newblock \bibinfo{title}{Majorana fermions in nuclear, particle, and
  solid-state physics}.
\newblock \emph{\bibinfo{journal}{Rev. Mod. Phys.}}
  \textbf{\bibinfo{volume}{87}}, \bibinfo{pages}{137--163} (\bibinfo{year}{2015}).

\bibitem{Inotani2012}
\bibinfo{author}{Inotani, D.}, \bibinfo{author}{Watanabe, R.},
  \bibinfo{author}{Sigrist, M.} \& \bibinfo{author}{Ohashi, Y.}
\newblock \bibinfo{title}{Pseudogap phenomenon in an ultracold {F}ermi gas with
  a $p$-wave pairing interaction}.
\newblock \emph{\bibinfo{journal}{Phys. Rev. A}} \textbf{\bibinfo{volume}{85}},
  \bibinfo{pages}{053628} (\bibinfo{year}{2012}).

\bibitem{Yoshida2015}
\bibinfo{author}{Yoshida, S.~M.} \& \bibinfo{author}{Ueda, M.}
\newblock \bibinfo{title}{{Universal High-Momentum Asymptote and Thermodynamic
  Relations in a Spinless Fermi Gas with a Resonant $p$-Wave Interaction}}.
\newblock \emph{\bibinfo{journal}{Phys. Rev. Lett.}}
  \textbf{\bibinfo{volume}{115}}, \bibinfo{pages}{135303}
  (\bibinfo{year}{2015}).

\bibitem{Yu2015}
\bibinfo{author}{Yu, Z.}, \bibinfo{author}{Thywissen, J.~H.} \&
  \bibinfo{author}{Zhang, S.}
\newblock \bibinfo{title}{{Universal Relations for a Fermi Gas Close to a
  $p$-Wave Interaction Resonance}}.
\newblock \emph{\bibinfo{journal}{Phys. Rev. Lett.}}
  \textbf{\bibinfo{volume}{115}}, \bibinfo{pages}{135304}
  (\bibinfo{year}{2015}).

\bibitem{Zhou:2016}
\bibinfo{author}{He, M.-Y.}, \bibinfo{author}{Zhang, S.-L.},
  \bibinfo{author}{Chan, H.~M.} \& \bibinfo{author}{Zhou, Q.}
\newblock \bibinfo{title}{{Concept of contact spectrum and its applications in
  atomic quantum Hall states}}.
\newblock \emph{\bibinfo{journal}{Phys. Rev. Lett.}}
  \textbf{\bibinfo{volume}{116}}, \bibinfo{pages}{045301}
  (\bibinfo{year}{2016}).

\bibitem{Zhang2010}
\bibinfo{author}{Zhang, P.}, \bibinfo{author}{Naidon, P.} \&
  \bibinfo{author}{Ueda, M.}
\newblock \bibinfo{title}{Scattering amplitude of ultracold atoms near the
  $p$-wave magnetic {F}eshbach resonance}.
\newblock \emph{\bibinfo{journal}{Phys. Rev. A}} \textbf{\bibinfo{volume}{82}},
  \bibinfo{pages}{062712} (\bibinfo{year}{2010}).

\bibitem{Ticknor2004}
\bibinfo{author}{Ticknor, C.}, \bibinfo{author}{Regal, C.~A.},
  \bibinfo{author}{Jin, D.~S.} \& \bibinfo{author}{Bohn, J.~L.}
\newblock \bibinfo{title}{Multiplet structure of {F}eshbach resonances in
  nonzero partial waves}.
\newblock \emph{\bibinfo{journal}{Phys. Rev. A}} \textbf{\bibinfo{volume}{69}},
  \bibinfo{pages}{042712} (\bibinfo{year}{2004}).

\bibitem{Chin2010}
\bibinfo{author}{Chin, C.}, \bibinfo{author}{Grimm, R.},
  \bibinfo{author}{Julienne, P.} \& \bibinfo{author}{Tiesinga, E.}
\newblock \bibinfo{title}{{F}eshbach resonances in ultracold gases}.
\newblock \emph{\bibinfo{journal}{Rev. Mod. Phys.}}
  \textbf{\bibinfo{volume}{82}}, \bibinfo{pages}{1225--1286}
  (\bibinfo{year}{2010}).

\bibitem{Jona2008}
\bibinfo{author}{Jona-Lasinio, M.}, \bibinfo{author}{Pricoupenko, L.} \&
  \bibinfo{author}{Castin, Y.}
\newblock \bibinfo{title}{Three fully polarized fermions close to a
  $\mathit{p}$-wave {F}eshbach resonance}.
\newblock \emph{\bibinfo{journal}{Phys. Rev. A}} \textbf{\bibinfo{volume}{77}},
  \bibinfo{pages}{043611} (\bibinfo{year}{2008}).

\bibitem{Gunter2005}
\bibinfo{author}{G\"unter, K.}, \bibinfo{author}{St\"oferle, T.},
  \bibinfo{author}{Moritz, H.}, \bibinfo{author}{K\"ohl, M.} \&
  \bibinfo{author}{Esslinger, T.}
\newblock \bibinfo{title}{$p$-wave interactions in low-dimensional fermionic
  gases}.
\newblock \emph{\bibinfo{journal}{Phys. Rev. Lett.}}
  \textbf{\bibinfo{volume}{95}}, \bibinfo{pages}{230401}
  (\bibinfo{year}{2005}).

\bibitem{Peng2014}
\bibinfo{author}{Peng, S.-G.}, \bibinfo{author}{Tan, S.} \&
  \bibinfo{author}{Jiang, K.}
\newblock \bibinfo{title}{Manipulation of $p$-wave scattering of cold atoms in
  low dimensions using the magnetic field vector}.
\newblock \emph{\bibinfo{journal}{Phys. Rev. Lett.}}
  \textbf{\bibinfo{volume}{112}}, \bibinfo{pages}{250401}
  (\bibinfo{year}{2014}).

\bibitem{Hazlett:2012dt}
\bibinfo{author}{Hazlett, E.~L.}, \bibinfo{author}{Zhang, Y.},
  \bibinfo{author}{Stites, R.~W.} \& \bibinfo{author}{O{\textquoteright}Hara,
  K.~M.}
\newblock \bibinfo{title}{{Realization of a resonant Fermi gas with a large
  effective range}}.
\newblock \emph{\bibinfo{journal}{Phys. Rev. Lett.}}
  \textbf{\bibinfo{volume}{108}}, \bibinfo{pages}{045304}
  (\bibinfo{year}{2012}).

\bibitem{Kohstall:2012kg}
\bibinfo{author}{Kohstall, C.}, \bibinfo{author}{Zaccanti, M.},
  \bibinfo{author}{Jag, M.} \& \bibinfo{author}{Trenkwalder, A.}
\newblock \bibinfo{title}{{Metastability and coherence of repulsive polarons in
  a strongly interacting Fermi mixture}}.
\newblock \emph{\bibinfo{journal}{Nature}} \textbf{\bibinfo{volume}{485}},
  \bibinfo{pages}{615--618} (\bibinfo{year}{2012}).

\bibitem{Gaebler2007}
\bibinfo{author}{Gaebler, J.~P.}, \bibinfo{author}{Stewart, J.~T.},
  \bibinfo{author}{Bohn, J.~L.} \& \bibinfo{author}{Jin, D.~S.}
\newblock \bibinfo{title}{$p$-wave {F}eshbach molecules}.
\newblock \emph{\bibinfo{journal}{Phys. Rev. Lett.}}
  \textbf{\bibinfo{volume}{98}}, \bibinfo{pages}{200403}
  (\bibinfo{year}{2007}).

\bibitem{Chin2005}
\bibinfo{author}{Chin, C.} \& \bibinfo{author}{Julienne, P.~S.}
\newblock \bibinfo{title}{Radio-frequency transitions on weakly bound ultracold
  molecules}.
\newblock \emph{\bibinfo{journal}{Phys. Rev. A}} \textbf{\bibinfo{volume}{71}},
  \bibinfo{pages}{012713} (\bibinfo{year}{2005}).

\bibitem{Pieri2009}
\bibinfo{author}{Pieri, P.}, \bibinfo{author}{Perali, A.} \&
  \bibinfo{author}{Strinati, G.~C.}
\newblock \bibinfo{title}{{Enhanced paraconductivity-like fluctuations in the
  radiofrequency spectra of ultracold {F}ermi atoms}}.
\newblock \emph{\bibinfo{journal}{Nat. Phys.}} \textbf{\bibinfo{volume}{5}},
  \bibinfo{pages}{736--740} (\bibinfo{year}{2009}).

\bibitem{Schneider2010}
\bibinfo{author}{Schneider, W.} \& \bibinfo{author}{Randeria, M.}
\newblock \bibinfo{title}{Universal short-distance structure of the
  single-particle spectral function of dilute {F}ermi gases}.
\newblock \emph{\bibinfo{journal}{Phys. Rev. A}} \textbf{\bibinfo{volume}{81}},
  \bibinfo{pages}{021601} (\bibinfo{year}{2010}).

\bibitem{Braaten2010}
\bibinfo{author}{Braaten, E.}, \bibinfo{author}{Kang, D.} \&
  \bibinfo{author}{Platter, L.}
\newblock \bibinfo{title}{Short-time operator product expansion for rf
  spectroscopy of a strongly interacting {F}ermi gas}.
\newblock \emph{\bibinfo{journal}{Phys. Rev. Lett.}}
  \textbf{\bibinfo{volume}{104}}, \bibinfo{pages}{223004}
  (\bibinfo{year}{2010}).

\bibitem{Stewart2010}
\bibinfo{author}{Stewart, J.~T.}, \bibinfo{author}{Gaebler, J.~P.},
  \bibinfo{author}{Drake, T.~E.} \& \bibinfo{author}{Jin, D.~S.}
\newblock \bibinfo{title}{Verification of universal relations in a strongly
  interacting {F}ermi gas}.
\newblock \emph{\bibinfo{journal}{Phys. Rev. Lett.}}
  \textbf{\bibinfo{volume}{104}}, \bibinfo{pages}{235301}
  (\bibinfo{year}{2010}).

\bibitem{Chevy2005}
\bibinfo{author}{Chevy, F.} \emph{et~al.}
\newblock \bibinfo{title}{Resonant scattering properties close to a $p$-wave
  {F}eshbach resonance}.
\newblock \emph{\bibinfo{journal}{Phys. Rev. A}} \textbf{\bibinfo{volume}{71}},
  \bibinfo{pages}{062710} (\bibinfo{year}{2005}).

\bibitem{Inada2008}
\bibinfo{author}{Inada, Y.} \emph{et~al.}
\newblock \bibinfo{title}{Collisional properties of $p$-wave {F}eshbach
  molecules}.
\newblock \emph{\bibinfo{journal}{Phys. Rev. Lett.}}
  \textbf{\bibinfo{volume}{101}}, \bibinfo{pages}{100401}
  (\bibinfo{year}{2008}).

\bibitem{Nakasuji2013}
\bibinfo{author}{Nakasuji, T.}, \bibinfo{author}{Yoshida, J.} \&
  \bibinfo{author}{Mukaiyama, T.}
\newblock \bibinfo{title}{Experimental determination of $p$-wave scattering
  parameters in ultracold $^{6}${Li} atoms}.
\newblock \emph{\bibinfo{journal}{Phys. Rev. A}} \textbf{\bibinfo{volume}{88}},
  \bibinfo{pages}{012710} (\bibinfo{year}{2013}).

\bibitem{Ohashi2005}
\bibinfo{author}{Ohashi, Y.}
\newblock \bibinfo{title}{{BCS-BEC} crossover in a gas of {F}ermi atoms with a
  $p$-wave {F}eshbach resonance}.
\newblock \emph{\bibinfo{journal}{Phys. Rev. Lett.}}
  \textbf{\bibinfo{volume}{94}}, \bibinfo{pages}{050403}
  (\bibinfo{year}{2005}).

\bibitem{Gurarie:2007gs}
\bibinfo{author}{Gurarie, V.} \& \bibinfo{author}{Radzihovsky, L.}
\newblock \bibinfo{title}{{Resonantly paired fermionic superfluids}}.
\newblock \emph{\bibinfo{journal}{Ann. Phys.}} \textbf{\bibinfo{volume}{322}},
  \bibinfo{pages}{2--119} (\bibinfo{year}{2007}).

\bibitem{Sagi2012}
\bibinfo{author}{Sagi, Y.}, \bibinfo{author}{Drake, T.~E.},
  \bibinfo{author}{Paudel, R.} \& \bibinfo{author}{Jin, D.~S.}
\newblock \bibinfo{title}{Measurement of the homogeneous contact of a unitary
  {F}ermi gas}.
\newblock \emph{\bibinfo{journal}{Phys. Rev. Lett.}}
  \textbf{\bibinfo{volume}{109}}, \bibinfo{pages}{220402}
  (\bibinfo{year}{2012}).

\bibitem{Pricoupenko2006}
\bibinfo{author}{Pricoupenko, L.}
\newblock \bibinfo{title}{Modeling interactions for resonant $p$-wave
  scattering}.
\newblock \emph{\bibinfo{journal}{Phys. Rev. Lett.}}
  \textbf{\bibinfo{volume}{96}}, \bibinfo{pages}{050401}
  (\bibinfo{year}{2006}).

\bibitem{Shenoy:2011dm}
\bibinfo{author}{Shenoy, V.~B.} \& \bibinfo{author}{Ho, T.-L.}
\newblock \bibinfo{title}{{Nature and Properties of a Repulsive Fermi Gas in
  the Upper Branch of the Energy Spectrum}}.
\newblock \emph{\bibinfo{journal}{Phys. Rev. Lett.}}
  \textbf{\bibinfo{volume}{107}}, \bibinfo{pages}{210401}
  (\bibinfo{year}{2011}).

\bibitem{Fuchs2008}
\bibinfo{author}{Fuchs, J.} \emph{et~al.}
\newblock \bibinfo{title}{{Binding energies of $^{6}${Li} $p$-wave {F}eshbach
  molecules}}.
\newblock \emph{\bibinfo{journal}{Phys. Rev. A}} \textbf{\bibinfo{volume}{77}},
  \bibinfo{pages}{053616} (\bibinfo{year}{2008}).

\bibitem{Gubbels2007}
\bibinfo{author}{Gubbels, K.~B.} \& \bibinfo{author}{Stoof, H. T.~C.}
\newblock \bibinfo{title}{Theory for $p$-wave {F}eshbach molecules}.
\newblock \emph{\bibinfo{journal}{Phys. Rev. Lett.}}
  \textbf{\bibinfo{volume}{99}}, \bibinfo{pages}{190406}
  (\bibinfo{year}{2007}).

\bibitem{Jin2008}
\bibinfo{author}{Jin, D.~S.}, \bibinfo{author}{Gaebler, J.~P.} \&
  \bibinfo{author}{Stewart, J.~T.}
\newblock \bibinfo{title}{An atomic {F}ermi gas near a p-wave {F}eshbach
  resonance}.
\newblock In \bibinfo{editor}{Hollberg, L.}, \bibinfo{editor}{Bergquist, J.} \&
  \bibinfo{editor}{Kasevich, M.} (eds.) \emph{\bibinfo{booktitle}{Proceedings
  of the XVIII International Conference on Laser Spectroscopy}},
  \bibinfo{pages}{127--137} (\bibinfo{publisher}{World Scientific},
  \bibinfo{address}{Singapore}, \bibinfo{year}{2008}).

\bibitem{Partridge2005}
\bibinfo{author}{Partridge, G.~B.}, \bibinfo{author}{Strecker, K.~E.},
  \bibinfo{author}{Kamar, R.~I.}, \bibinfo{author}{Jack, M.~W.} \&
  \bibinfo{author}{Hulet, R.~G.}
\newblock \bibinfo{title}{Molecular probe of pairing in the {BEC-BCS}
  crossover}.
\newblock \emph{\bibinfo{journal}{Phys. Rev. Lett.}}
  \textbf{\bibinfo{volume}{95}}, \bibinfo{pages}{020404}
  (\bibinfo{year}{2005}).

\bibitem{Kuhnle2010}
\bibinfo{author}{Kuhnle, E.~D.} \emph{et~al.}
\newblock \bibinfo{title}{Universal behavior of pair correlations in a strongly
  interacting {F}ermi gas}.
\newblock \emph{\bibinfo{journal}{Phys. Rev. Lett.}}
  \textbf{\bibinfo{volume}{105}}, \bibinfo{pages}{070402}
  (\bibinfo{year}{2010}).

\bibitem{Navon:2010ix}
\bibinfo{author}{Navon, N.}, \bibinfo{author}{Nascimb{\`e}ne, S.},
  \bibinfo{author}{Chevy, F.} \& \bibinfo{author}{Salomon, C.}
\newblock \bibinfo{title}{{The Equation of State of a Low-Temperature {F}ermi
  Gas with Tunable Interactions}}.
\newblock \emph{\bibinfo{journal}{Science}} \textbf{\bibinfo{volume}{328}},
  \bibinfo{pages}{729--732} (\bibinfo{year}{2010}).

\bibitem{Kuhnle2011}
\bibinfo{author}{Kuhnle, E.~D.} \emph{et~al.}
\newblock \bibinfo{title}{Temperature dependence of the universal contact
  parameter in a unitary {F}ermi gas}.
\newblock \emph{\bibinfo{journal}{Phys. Rev. Lett.}}
  \textbf{\bibinfo{volume}{106}}, \bibinfo{pages}{170402}
  (\bibinfo{year}{2011}).

\bibitem{Levinsen2008}
\bibinfo{author}{Levinsen, J.}, \bibinfo{author}{Cooper, N.~R.} \&
  \bibinfo{author}{Gurarie, V.}
\newblock \bibinfo{title}{Stability of fermionic gases close to a $p$-wave
  {F}eshbach resonance}.
\newblock \emph{\bibinfo{journal}{Phys. Rev. A}} \textbf{\bibinfo{volume}{78}},
  \bibinfo{pages}{063616} (\bibinfo{year}{2008}).

\bibitem{Gurarie:2005it}
\bibinfo{author}{Gurarie, V.}, \bibinfo{author}{Radzihovsky, L.} \&
  \bibinfo{author}{Andreev, A.~V.}
\newblock \bibinfo{title}{{Quantum Phase Transitions across a p-Wave Feshbach
  Resonance}}.
\newblock \emph{\bibinfo{journal}{Phys. Rev. Lett.}}
  \textbf{\bibinfo{volume}{94}}, \bibinfo{pages}{230403}
  (\bibinfo{year}{2005}).

\bibitem{Cheng:2005kv}
\bibinfo{author}{Cheng, C.-H.} \& \bibinfo{author}{Yip, S.~K.}
\newblock \bibinfo{title}{{Anisotropic Fermi Superfluid via p-Wave Feshbach
  Resonance}}.
\newblock \emph{\bibinfo{journal}{Phys. Rev. Lett.}}
  \textbf{\bibinfo{volume}{95}}, \bibinfo{pages}{070404}
  (\bibinfo{year}{2005}).
  
\end{thebibliography}
\end{document}